\documentclass[twocolumn,showpacs,preprintnumbers,amsmath,amssymb,nofootinbib]{revtex4}

\usepackage{ifthen}
\usepackage{hyperref}
\usepackage{graphicx}
\usepackage{axodraw}

\newboolean{showPurple}
\setboolean{showPurple}{false}

\preprint{DAMTP-2012-40, 
TIFR/TH/12-20}

\ifthenelse{\boolean{showPurple}}{\newcommand{\showPur}[1]{#1}}
{\newcommand{\showPur}[1]{}}

\begin{document}

\title{$R-$Parity Violating Supersymmetry Explanation for the Large $t \bar t$
Forward-Backward Asymmetry} 
\date{\today}
\author{B.C. Allanach}
\affiliation{DAMTP, CMS, University of Cambridge, Wilberforce Road, Cambridge, CB3 0WA, United Kingdom}
\author{K. Sridhar}
\affiliation{TIFR, Homi Bhabha Road, Colaba, Mumbai 400 005, India}

\begin{abstract}
We examine a supersymmetric explanation for 
the anomalously high forward backward asymmetry in top pair production
measured by CDF and D0. We suppose that it is due to the $t-$channel
exchange of a right-handed sbottom which couples to $d_R$ and $t_R$, as is
present 
in the $R-$parity violating minimal supersymmetric standard model.
We show that all Tevatron and LHC experiments'
$t \bar t$ constraints
may be respected for a
sbottom mass between 300 and 1200 GeV, and a large Yukawa coupling $>2.2$,
yielding $A_{FB}$ up to 0.18. The
non Standard Model contribution to the 
LHC charge asymmetry parameter is $\Delta A_C^y=0.017-0.045$,
small 
enough to 
be consistent with current measurements but non-zero and positive,
allowing for LHC confirmation in the future within 20fb$^{-1}$. A small
additional contribution 
to the LHC $t \bar t$ production cross-section is also predicted, allowing a
further test. We estimate that 10 fb$^{-1}$ of LHC luminosity would be
sufficient to  rule out 
the proposal to 95$\%$ confidence level, if the measurements of the $t \bar t$
cross-section turn out to be centred on the Standard Model prediction.
\end{abstract}

\maketitle


In the 1.96 TeV centre of mass energy $p \bar p$ collisions  at the
Tevatron collider, measurements of the $t \bar t$ forward
backward asymmetry $A_{FB}$ 
were made by the CDF and D0 experiments. A positive non-zero value indicates
that, in $t \bar t$ production events, 
a higher number of events\footnote{$c=\cos \theta$, where $\theta$ is defined to be the scattering angle
  between incoming proton beam and outgoing top in the centre of mass frame of
  $t \bar t$.} $N(c>0)$ had a $t$
travelling in a more forward direction than the $\bar t$ compared
to the number $N(c
<0)$ of those events where the $t$ travelled in a more backward direction than the 
$\bar t$ 
\begin{equation}
A_{FB} = \frac{N(c>0) - N(c<0)}{N(c>0) + N(c< 0)}
\end{equation}
The Standard Model (SM) prediction of this quantity is $A_{FB}^{SM}=0.066 \pm
0.020$~\cite{Ahrens:2011uf}, 
and dominantly derives from the interference between tree-level and one-loop quantum chromodynamics (QCD)
diagrams. CDF measured an unfolded
value\footnote{Different shower models can produce different values when they
  are  used in the
  unfolding because of the different treatments of QCD
  coherence~\cite{Skands:2012mm}, yielding effects on $A_{FB}$
  of order several percent.}
0.158$\pm$0.075~\cite{PhysRevD.83.112003,10584}, 
whereas D0 measured 
0.196$\pm$0.065~\cite{PhysRevD.84.112005}, each significantly higher 
than $A_{FB}^{SM}$. 
We suppose here that the discrepancy between measurements and the SM prediction
for $A_{FB}$ is due to a particular
beyond the SM process, and examine other constraints to see if the explanation
remains viable.

Since the LHC is a $pp$ collider and thus has an initial state which is
symmetric under $c\leftrightarrow -c$, $A_{FB}^{LHC}=0$. 
However, the LHC is able to measure a related but different charge asymmetry
in the  number of tops that are travelling at a smaller angle to the beam-line
compared to the number of anti-tops that are travelling closer to the
beam-line: 
\begin{equation}
A_C^y=\frac{N(|y_t|>|y_{\bar t}|)-N(|y_{\bar t}|>|y_t|)}
{N(|y_t|>|y_{\bar t}|)+N(|y_{\bar t}|>|y_t|)},
\end{equation}
where $y_i=1/2 \ln(E_i+{p_i}_z)/(E_i-{p_i}_z)$ is the rapidity of particle
$i$.
The SM prediction for 7 TeV collisions is ${A_C^y}^{SM}=0.006 \pm 0.002$,
which is 
consistent with the combined ATLAS and CMS
measurements $A_C^y=-0.015 \pm 0.04$~\cite{ATLAS:2012an,Chatrchyan:2011hk}. Any 
non-SM explanation for the high measured value of $A_{FB}$ must also
therefore not predict too high a value for a non-SM contribution $\Delta
A_C^y=A_C^y-{A_C^y}^{SM}$.

Many models have so far been proposed to explain $A_{FB}$ measurements and
several have
failed other constraints. Axigluons, $W'$ and $Z'$ vector
bosons, as well as the $t-$channel exchange of various
scalars~\cite{Shu:2009xf,AguilarSaavedra:2012ma,Blum:2011fa} have all been
proposed.   
Here, we show that the exchange of a particular scalar: a right-handed
sbottom, can explain the apparent sizable enhancement to $A_{FB}$ while
respecting other relevant constraints. 
Ref.~\cite{Cao:2009uz} included this, among other possibilities,
as an explanation for $A_{FB}$. However, the authors only considered
couplings smaller than 1.25 because they required perturbativity 
up until the GUT scale, and found that the new physics contributions to 
$A_{FB}$ were too small to obtain it to within the 1$\sigma$ measurement
errors, although they {\em were}\/ able to obtain it to just within 2$\sigma$
measurement errors.   
We shall consider
larger values of the coupling, which we show are necessary to explain the data.
Our work also goes further than Ref.~\cite{Cao:2009uz} in
the sense that we consider the LHC charge asymmetry and top production
constraints as well.

\begin{figure}
 \begin{center}
\includegraphics[width=4cm]{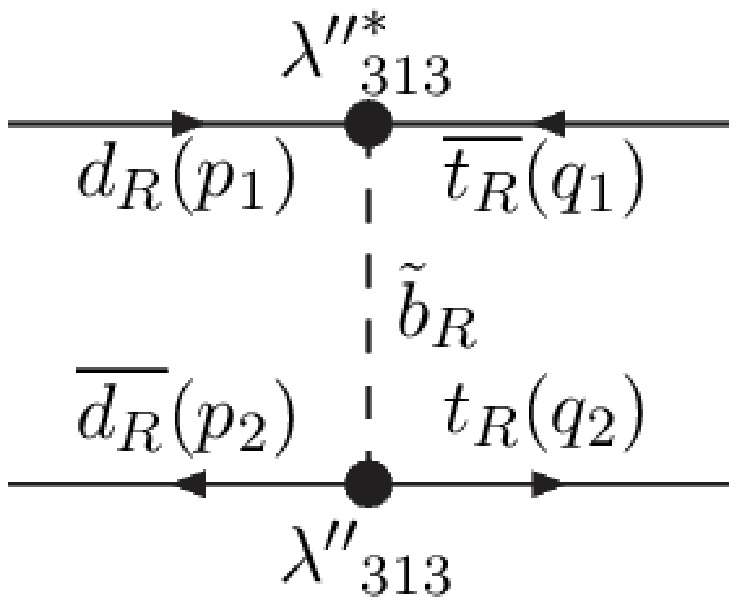}
 \caption{Leading order supersymmetric (SUSY) contribution to $A_{FB}$ from the
   RPV coupling $\lambda''_{313}$ and right-handed sbottom
   exchange. $p_1, p_2, q_1$ and $q_2$ denote 4-momenta.\label{fig:LOprocess}} 
 \end{center}
 \end{figure}

The $R-$parity violating (RPV) interactions of the minimal supersymmetric
standard 
model (MSSM) include the superpotential
term 
\begin{equation}
W=\frac{\lambda''_{313}}{2} {\bar T_R} {\bar D_R} {\bar B_R},
\label{coupling}
\end{equation}
where gauge indices have been suppressed and $\bar T_R$, $\bar D_R$, $\bar
B_R$ are chiral superfields containing the anti-right handed top $\overline{
  t_R}$, 
anti-right 
handed down $\overline{d_R}$ and anti-right handed bottom quark
$\overline{b_R}$, 
respectively. $W=\lambda''_{312}{\bar T_R} {\bar
  D_R} {\bar S_R}$ could also explain $A_{FB}$, although the constraints on 
$m_{\tilde s_R}$ are likely to be stronger than those on the $\tilde b_R$, since
the strange PDFs are higher than the bottom PDFs, and so they will be
predicted to be produced more readily at the LHC\@. 
The operator in Eq.~\ref{coupling} has an
antisymmetric colour structure. 
It leads to an additional tree-level process
that contributes to $A_{FB}$, the Feynman diagram of which is shown in
Fig.~\ref{fig:LOprocess}.
Ref.~\cite{Cao:2009uz} put an upper bound on $\lambda''_{313}<1.25$ on the
grounds of 
perturbativity up to the GUT scale. In the present paper, we shall not worry
about 
a premature ultra-violet completion of our model: the coupling will reach a
Landau pole 
around 10-100 TeV or so, and above that scale the effective theory will then
change.  
We shall consider
couplings as large as 5, although we note that at the higher end above
3.5, our predictions may start to become less accurate due to larger higher
loop corrections that we do not take into account.

The tree-level $\lambda''_{313} \neq 0$ contribution to the differential
cross-section of $t \bar t$ 
production is, where $\Gamma = | \lambda''_{313}|^2 m_{{\tilde b}_R}(1-m_t^2/m_{{\tilde b}_R}^2)^2
/ (8 \pi)$ is the sbottom width,
\begin{eqnarray}
\frac{d \Delta \sigma}{d c} &=& 
\frac{|\lambda''_{313}|^4 \beta \hat s}{96 \pi }  \frac{( \beta c+1)^2}
{[\hat s(\beta c +1)+ 2m_{{\tilde b}_R}^2-2m_t^2]^2+4m_{{\tilde b}_R}^2
  \Gamma^2} \nonumber \\ 
&+&\frac{\alpha_s |\lambda''_{313}|^2 \beta }{36 \hat s} 
[\hat s (1 + \beta c) + 2 m_t^2 - 2
  m_{{\tilde b}_R}^2]
 \nonumber \\
&\times& \frac{\hat s(1+\beta c)^2+4 m_t^2 }{[\hat s(\beta c +1)+ 2m_{{\tilde
      b}_R}^2
-2m_t^2]^2+4m_{{\tilde b}_R}^2
  \Gamma^2},\label{dsdcost}
\end{eqnarray}
where $\beta=\sqrt{1 - 4 m_t^2 / \hat s}$,
$\hat s=(p_1+p_2)^2$, 
$\alpha_s$ is the strong coupling constant
and $m_t$ is the top
quark mass. Eq.~\ref{dsdcost} predicts a
non-zero  
contribution to $A_{FB}$ since it is not even in $c$. It disagrees with the
expression for $d \Delta 
\sigma/d \hat t$ in Ref.~\cite{Ghosh:1996bm} by a factor of 64 in the first
term, but agrees with the recent Ref.~\cite{Dupuis:2012is}.
In order to calculate
the non-SM contribution to the asymmetry $\Delta A_{FB}$, we must convolute
the differential cross-section with the parton distribution functions (PDFs)
for (anti-)down quarks in (anti-)protons numerically. 
We have numerically checked Eq.~\ref{dsdcost} at a test point for large
$\lambda''_{313}$ (where the first term dominates) that integrating
Eq.~\ref{dsdcost} over the PDFs gives an identical result to the program {\tt
  MadGraph5\_v1\_4\_5}~\cite{Alwall:2011uj}. 

Of eight scalar models considered in Ref.~\cite{Blum:2011fa} that might
have explained the $A_{FB}$ measurements, the
exchange of a charge $-1/3$ colour triplet scalar that has a large
coupling to $d_R  \bar t_R$ was considered and discarded. 
As far as $A_{FB}$ goes, this is identical to our SUSY model
except for the other interactions of the MSSM (which are 
important for passing the atomic parity violation constraints, as mentioned
below). 
It was deduced that for this scalar, 
there is no parameter space that
simultaneously satisfies other constraints
as well as $\Delta A_{FB}^h>0.2$, where $^h$ implies the high invariant mass
constraint $m_{t \bar t}>450\mbox{~GeV}$. However, CDF data 
on $\Delta A_{FB}^h$ has lowered considerably since Ref.~\cite{Blum:2011fa},
with 
a new CDF 
measurement~\cite{moriond} implying $\Delta A_{FB}^h=0.20 \pm 0.07$. We shall
show that the model now agrees with all current relevant measurements. 
The measured value $A_{FB}^l=-0.116 \pm 0.153$, which is the asymmetry
with the additional cut $m_{t \bar t}<450\mbox{~GeV}$, is
consistent with the Standard Model prediction ${A_{FB}^l}^{SM}=0.04$.

We shall work in a `bottom-up' framework,
where we deal with weak-scale SUSY, making no assumptions about the
ultra-violet limit of our model (in particular, we do not demand
perturbativity of couplings up to a grand unified scale).
Thus, for the present, we shall work in a simplified model where we set all
sparticles of the MSSM to be heavy, aside from the right-handed
sbottom $\tilde b_R$, unless for some particular reason we require another
sparticle to play a r\^{o}le. 
The coupling $\lambda''_{313}$, if it is the
only non-negligible real RPV coupling, is not constrained to be
small by indirect experimental data~\cite{Allanach:1999ic,Barbier:2004ez} for 
$\tilde b_R$s that are not too light\footnote{If we were to have light
  charginos and left-handed down squarks, as 
well as a mixing between left and right-handed down squarks, we would have 
severe constraints upon $\lambda''_{313}$ coming from a loop diagram inducing
neutrino anti-neutron oscillations~\cite{Chang:1996sw}.
For instance, if all
sparticles were to have a mass of 200 GeV (and the left and right-handed
squarks were to be
mixed with a trilinear scalar coupling of 200 GeV), then the upper bound would
be $\lambda''_{313} < O(0.04)$.}.
The strongest constraints 
from $R_l=\Gamma(Z^0 \rightarrow \mbox{hadrons})/\Gamma(Z^0\rightarrow l
 \bar l)$~\cite{Bhattacharyya:1995bw} are still too weak (for the right-handed
 sbottom  masses $>300$ GeV that we shall be interested in) to be limiting.
Eq.~\ref{coupling} has exactly the right properties to evade current stringent
flavor constraints on di-quarks: it only couples to right-handed quarks and
induces no tree-level flavour changing neutral currents~\cite{Giudice:2011ak}.
Entertaining the possibility of other baryon-number violating couplings
$\lambda''_{ijk}$ 
in the presence of a large coupling $\lambda''_{313}$, there is a
particularly strict bound on $|\lambda''_{313} {\lambda''_{323}}^*|<
0.01$~\cite{Slavich:2000xm} coming from $K^0 - \bar K^0$ 
mixing constraints for sparticle mass input parameters less than 1
TeV. Thus, 
strong constraints upon  $|\lambda''_{323}|$ would 
apply, and some fermion mass model building would have to be performed to see
if this were realistic. It has been argued that simple Higgsed
abelian flavour symmetries and the Froggatt-Nielsen mechanism would naively
predict that the order of magnitude of $|\lambda''_{323}|$ would be larger
than that of $|\lambda''_{313}|$~\cite{Slavich:2000xm}. However, this order of
magnitude estimate is rather rough, and assumes that several couplings
involved in the Froggatt-Nielsen mechanism are of order 1, an assumption that
may be violated in more explicit constructions. Here, it seems premature to
apply any such model building, since as we shall show, LHC charge asymmetry
and top production cross-section measurements can
test our proposal. We thus proceed on the basis that $|\lambda''_{323}|$ is
small enough to evade this bound, approximating it by zero in our numerical
analysis. 

We now discuss the various constraints that we shall employ.
Naively adding errors in quadrature, we find a CDF and D0 weighted average of 
$A_{FB}=0.187\pm 0.037$. When combined with the SM prediction,
this implies a measured non-SM contribution of 
$\Delta A_{FB}=A_{FB}-A_{FB}^{SM}=0.121 \pm 0.042$.
On the other hand, Tevatron measurements of the SM total $t \bar t$ production
cross-section are roughly in line with SM predictions. A non-SM component is
restricted by~\cite{9913}
$\Delta \sigma^{TEV}_{t \bar t} = 0.43 \pm 0.54$ pb.
The differential production cross-section $d \sigma_{t \bar t}^{TEV}/
d m_{t  \bar t}$ 
was measured by CDF in Ref.~\cite{Aaltonen:2009iz}. We shall employ (following
Ref.~\cite{Blum:2011fa}) the
measurement 
$\sigma_{t \bar t}^{TEV}(700 \mbox{~GeV}<m_{t \bar t}<800 \mbox{~GeV})= 80 \pm
37 \mbox{~fb} $, 
versus a SM prediction of~\cite{Almeida:2008ug,Ahrens:2010zv} 80$\pm$8 pb, so 
the non-SM contribution $\Delta \sigma_{t \bar t}^{TEV}(\mbox{bin})$ must not be too
large. 
This invariant mass bin is far away from the bulk of the $t \bar t$
differential cross-section, and so ought to provide 
information that is approximately independent of the information from 
$\sigma^{TEV}_{t \bar t}$.  
At the 95$\%$ confidence level (CL), this leaves little room for 
a non-SM contribution of
$\Delta \sigma_{t \bar t}^{TEV}$(bin).
ATLAS and CMS have~\cite{atlas1,cms1}  measured the 7 TeV $pp \rightarrow t
\bar t$ cross-section to be 
$\sigma^{LHC7}_{t \bar t}=173.4 \pm 10.6$ pb, versus a SM
prediction of $163 \pm 10$ pb.  
In Table~\ref{tab:consts}, we summarise the constraints that we require
predicted observables to satisfy.
\begin{table}
\begin{tabular}{c|c} \hline
$0.037<\Delta A_{FB}<0.205$ & $-0.079<\Delta A_C^y<0.061$ \\
$-0.65<\Delta \sigma^{TEV}_{t \bar t}/\mbox{pb} <1.51 $ &
$-76<\Delta \sigma^{TEV}_{t \bar t}(\mbox{bin})$/fb$<76$ \\ 
$-0.38 < \Delta A_{FB}^l < 0.23$ &
$0.062 < \Delta A_{FB}^h < 0.33$ \\
$-19.2<\Delta \sigma_{t \bar t}^{LHC7}/\mbox{pb} < 39.2$ & 
\\
\hline
\end{tabular}
\caption{95$\%$ CL constraints on new physics contributions to observables
  that are brought 
  to bear upon our model. The 
  limits have been derived by using naive summation in quadrature of all errors.
\label{tab:consts}}
\end{table}

\begin{figure*}
\begin{center}
\begin{picture}(420,380)(0,-20)
\put(30,200){\begin{picture}(100,202)(55,0)
\put(-55,202){\includegraphics[angle=270,width=11.8cm]{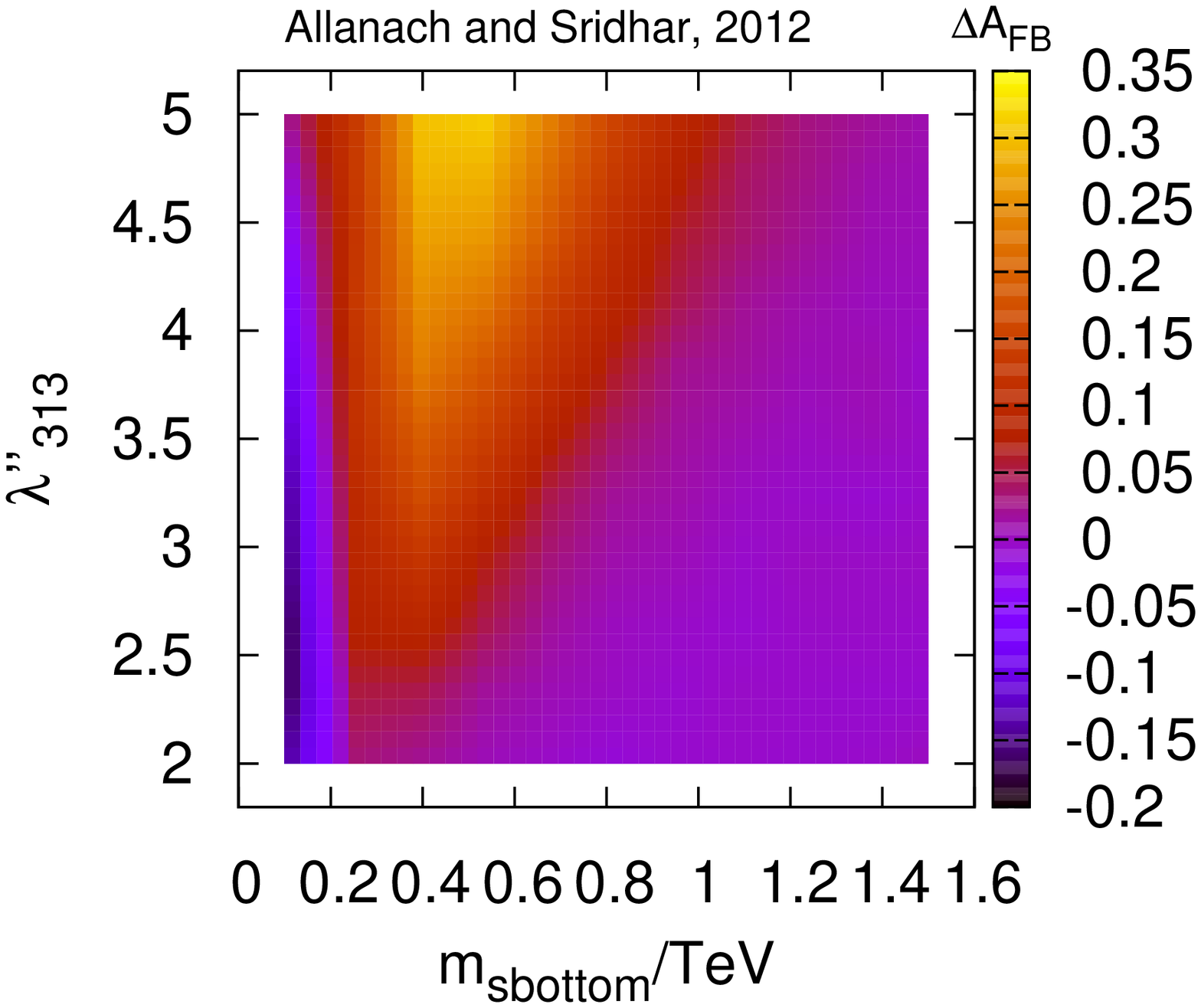}}
\put(0,170){\includegraphics[angle=270,width=8cm]{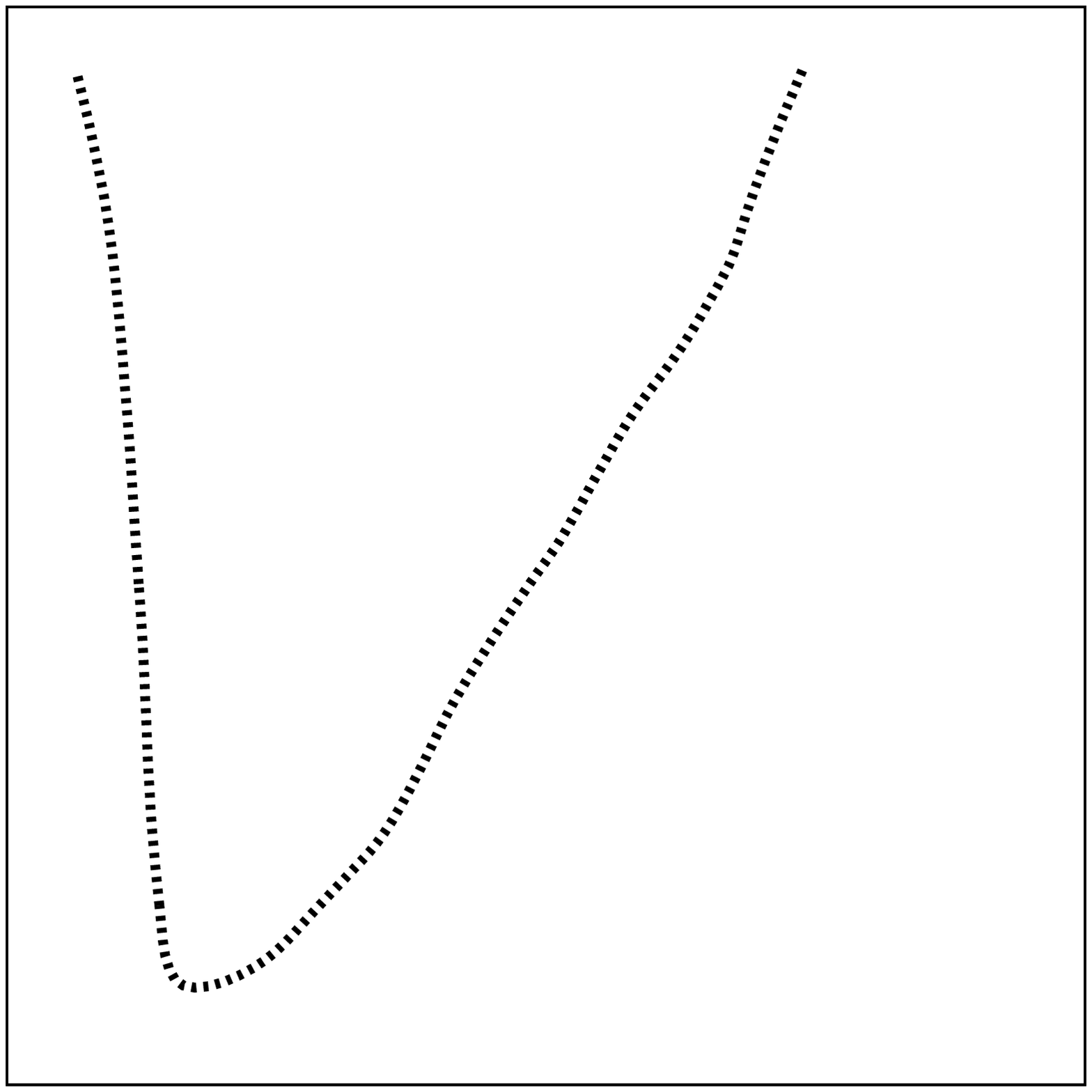}}
\put(0,170){\includegraphics[angle=270,width=8cm]{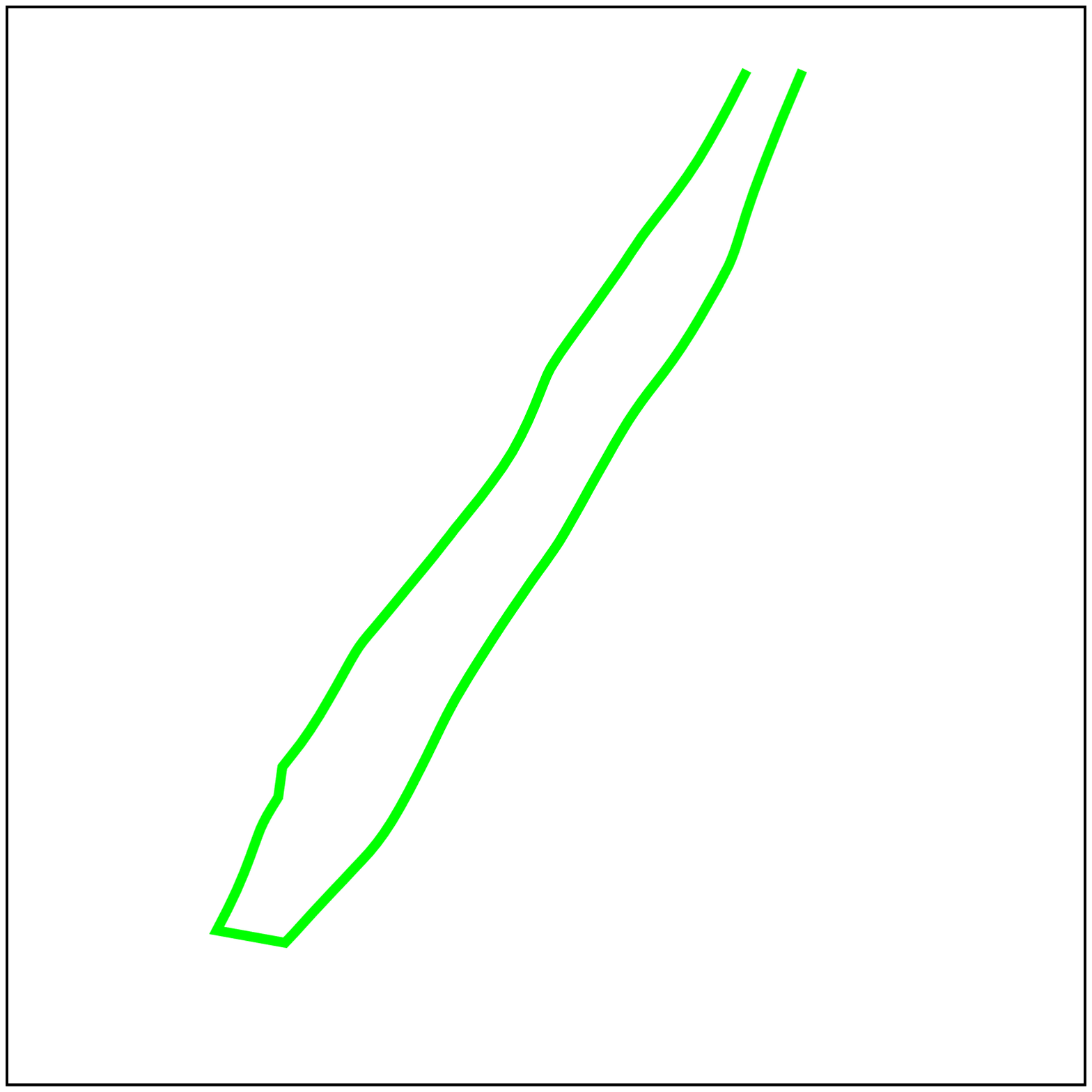}}
\showPur{\put(0,170){\includegraphics[angle=270,width=8cm]{valAfb.eps}}}
\end{picture}
}
\put(260,200){\begin{picture}(100,202)(55,0)
\put(-55,202){\includegraphics[angle=270,width=11.8cm]{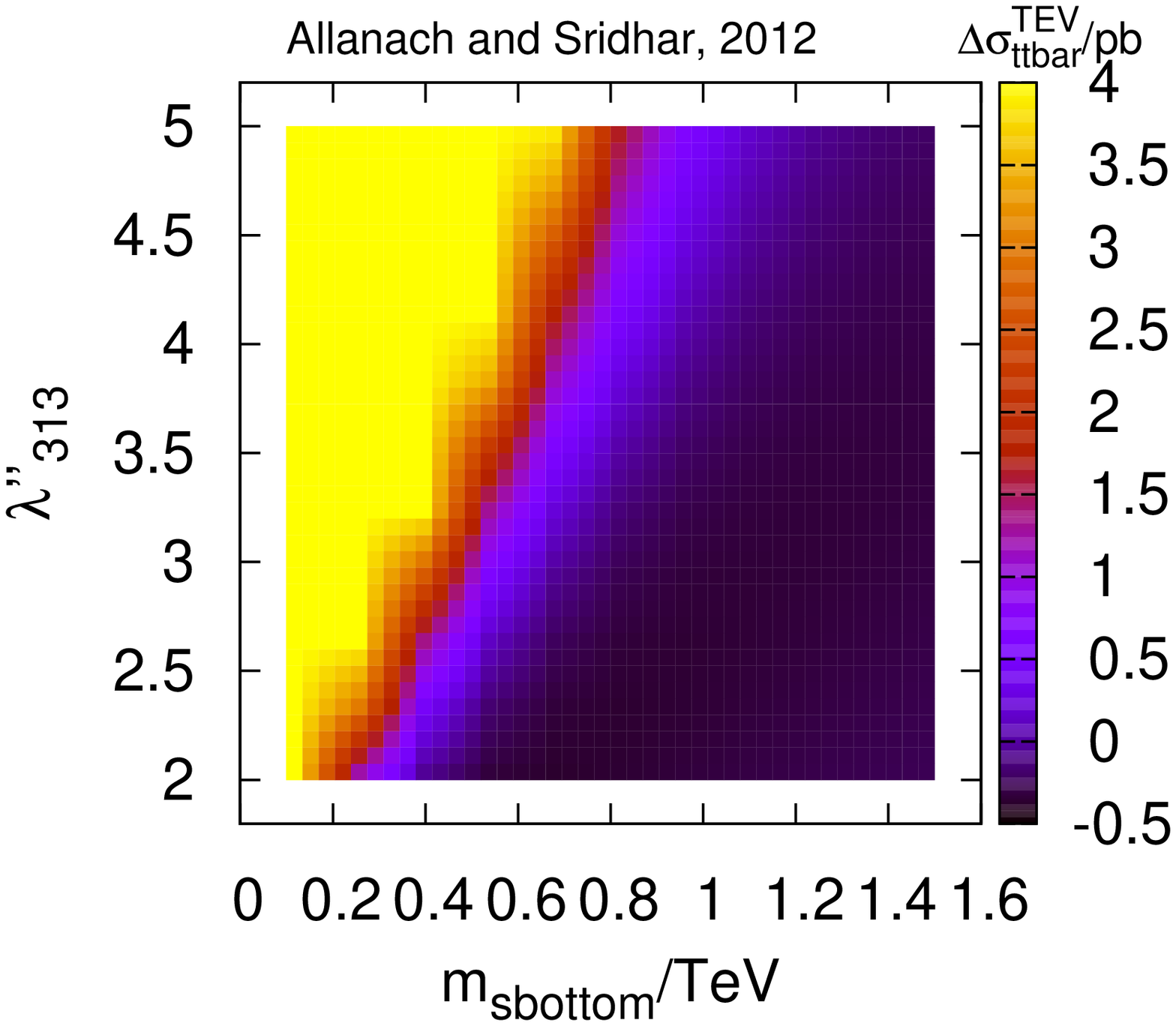}}
\put(0,170){\includegraphics[angle=270,width=8cm]{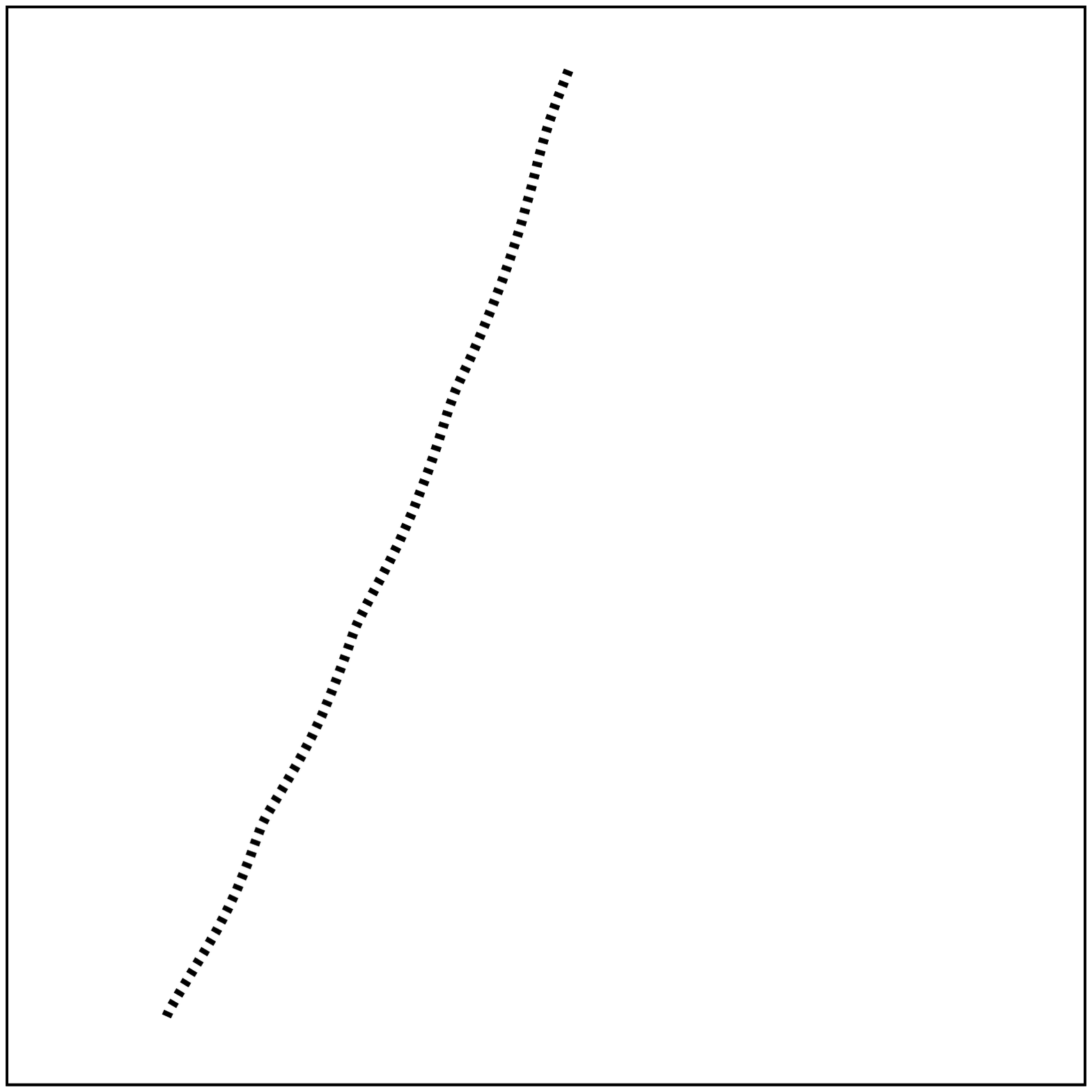}}
\put(0,170){\includegraphics[angle=270,width=8cm]{comb.eps}}
\showPur{\put(0,170){\includegraphics[angle=270,width=8cm]{valSigTev.eps}}}
\end{picture}
}
\put(30,0){\begin{picture}(100,202)(55,0)
\put(-55,202){\includegraphics[angle=270,width=11.8cm]{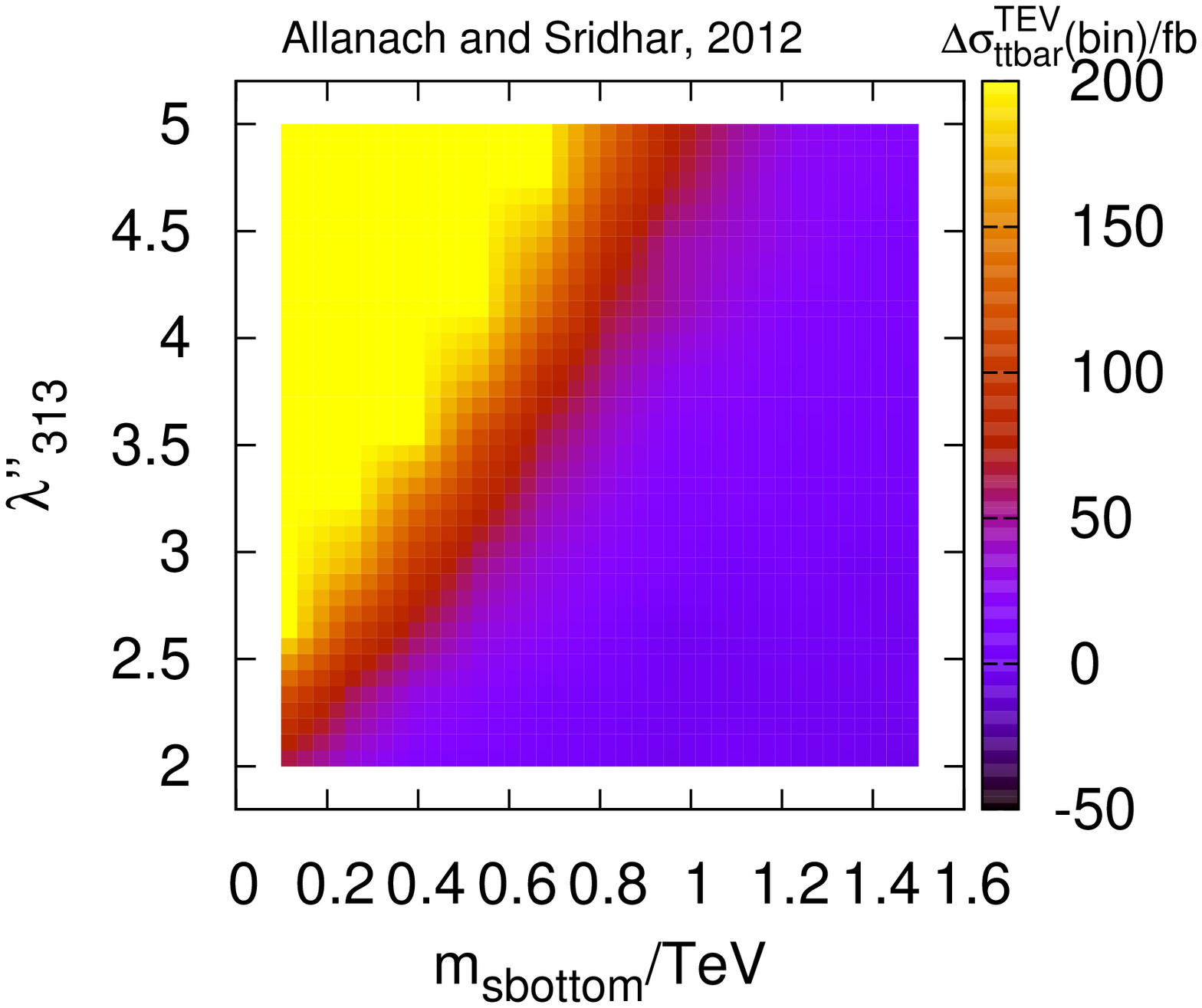}}
\put(0,170){\includegraphics[angle=270,width=8cm]{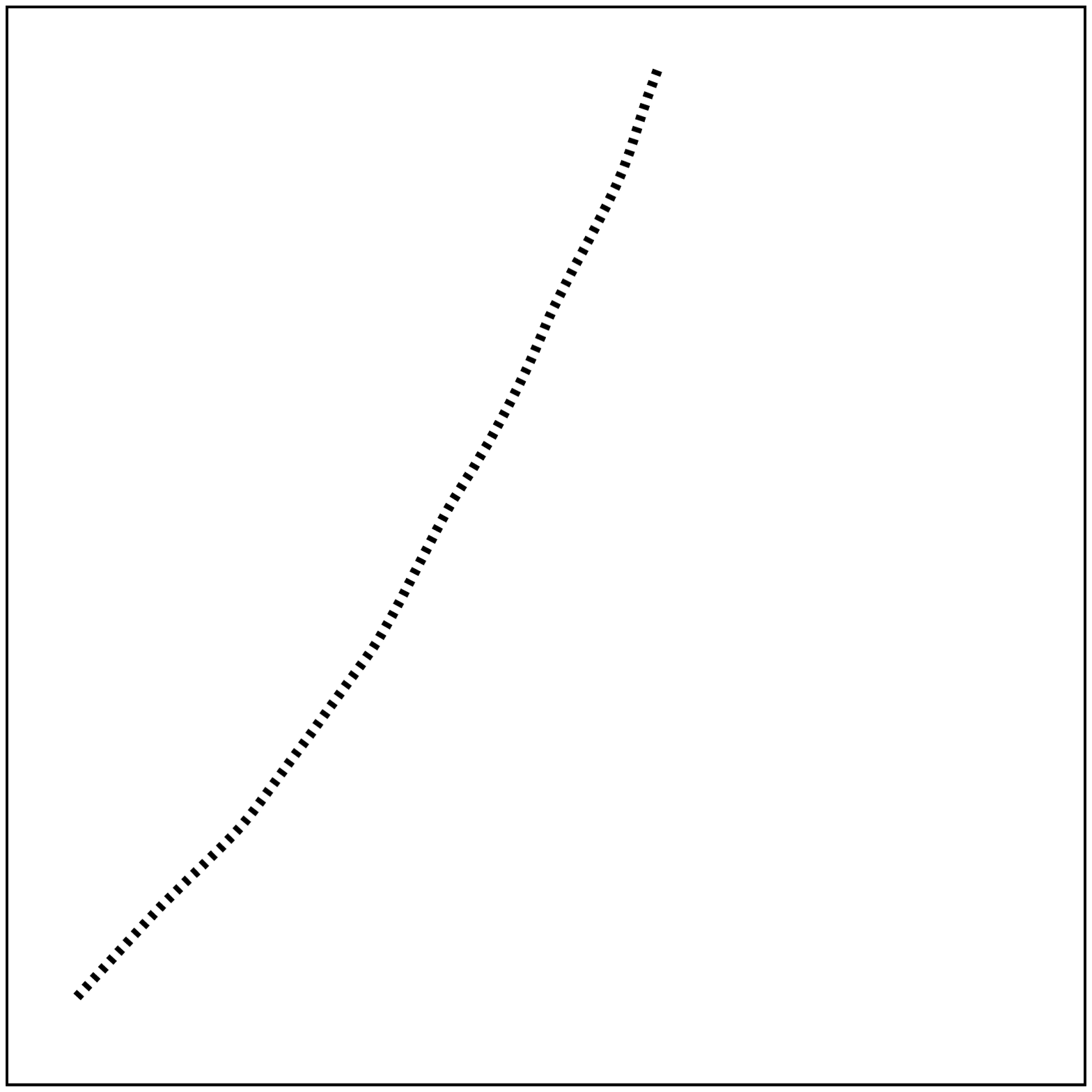}}
\put(0,170){\includegraphics[angle=270,width=8cm]{comb.eps}}
\showPur{\put(0,170){\includegraphics[angle=270,width=8cm]{valSigm.eps}}}
\end{picture}
}
\put(260,0){\begin{picture}(100,202)(55,0)
\put(-55,202){\includegraphics[angle=270,width=11.8cm]{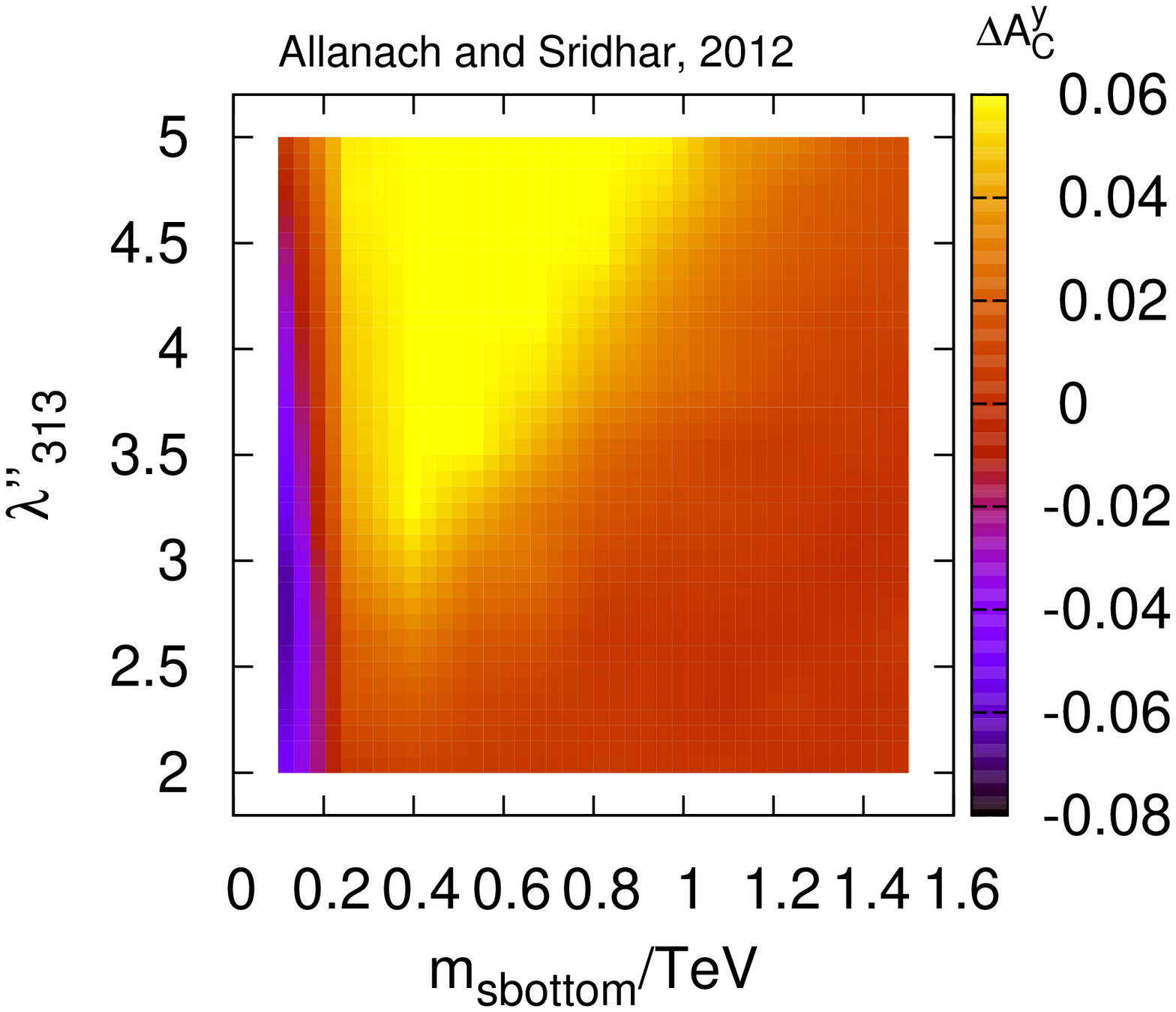}}
\put(0,170){\includegraphics[angle=270,width=8cm]{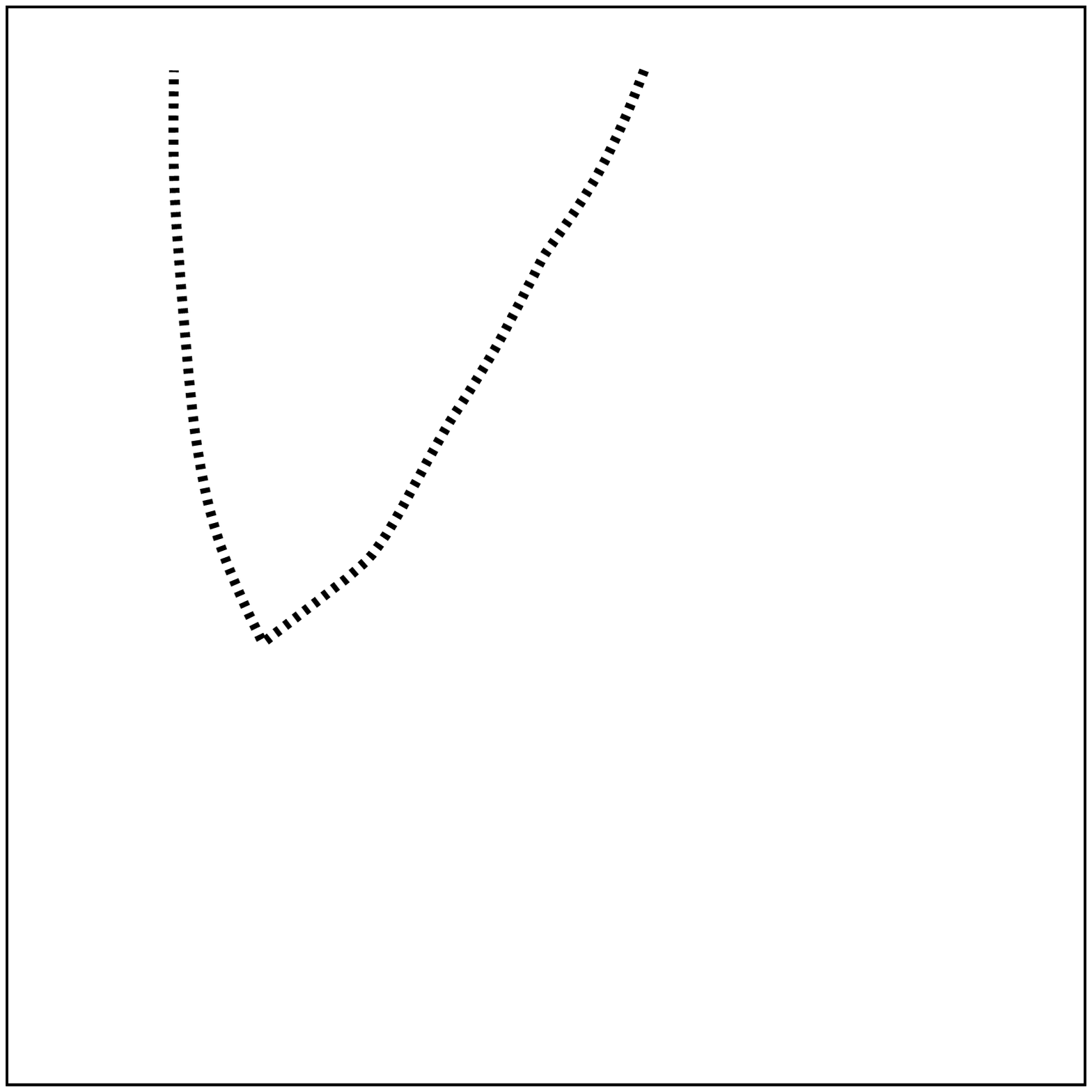}}
\put(0,170){\includegraphics[angle=270,width=8cm]{comb.eps}}
\showPur{\put(0,170){\includegraphics[angle=270,width=8cm]{valDy.eps}}}
\end{picture}
}
    \put(0,370){(a)}
    \put(230,370){(b)}
    \put(0,175){(c)}
    \put(230,175){(d)}
\end{picture}
\caption{Predicted non-SM contributions to various constraining
  observables: (a) the non-SM $t \bar t$ forward backward asymmetry parameter,
  (b) the total $t \bar t$ Tevatron cross-section, (c) the total $t \bar t$
  Tevatron cross-section in the bin $700$ GeV$<m_{t \bar t}<$800 GeV, (d) 
  the LHC charge asymmetry. Each individual constraint is respected below the
  broken contour, except for in (a), where it is respected above the contour.
  Inside the solid  contour, 
  all limits in Table~\protect\ref{tab:consts} are respected.
We have checked that the 8 TeV LHC has the same predictions for $A_C^y$ as those
displayed in (d), to a very good approximation.
\label{fig:afb}} 
\end{center}
\end{figure*}

Throughout the present paper, we calculate experimental observables using the
matrix element event generator {\tt
  MadGraph5\_v1\_1\_4\_5}~\cite{Alwall:2011uj} assuming $m_t=173.1$ GeV, the {\tt
 CTEQ6L1} PDFs~\cite{Kretzer:2003it} and using the {\tt
  FeynRules}~\cite{Christensen:2008py} implementation of the RPV
MSSM~\cite{Fuks:2012im,Duhr:2011se}. We define an 11 by 11 grid in $m_{{\tilde
    b}_R}$-$\lambda''_{313}$ parameter space,
simulating 100000 $t \bar t$ production events and interpolating predicted
observables in between the grid points. 

\showPur{\begin{figure*}
\begin{center}
\begin{picture}(420,400)(0,-20)
\put(30,200){\begin{picture}(100,202)(55,0)
\put(-55,202){\includegraphics[angle=270,width=11.8cm]{mlAfblow.eps}}
\put(0,170){\includegraphics[angle=270,width=8cm]{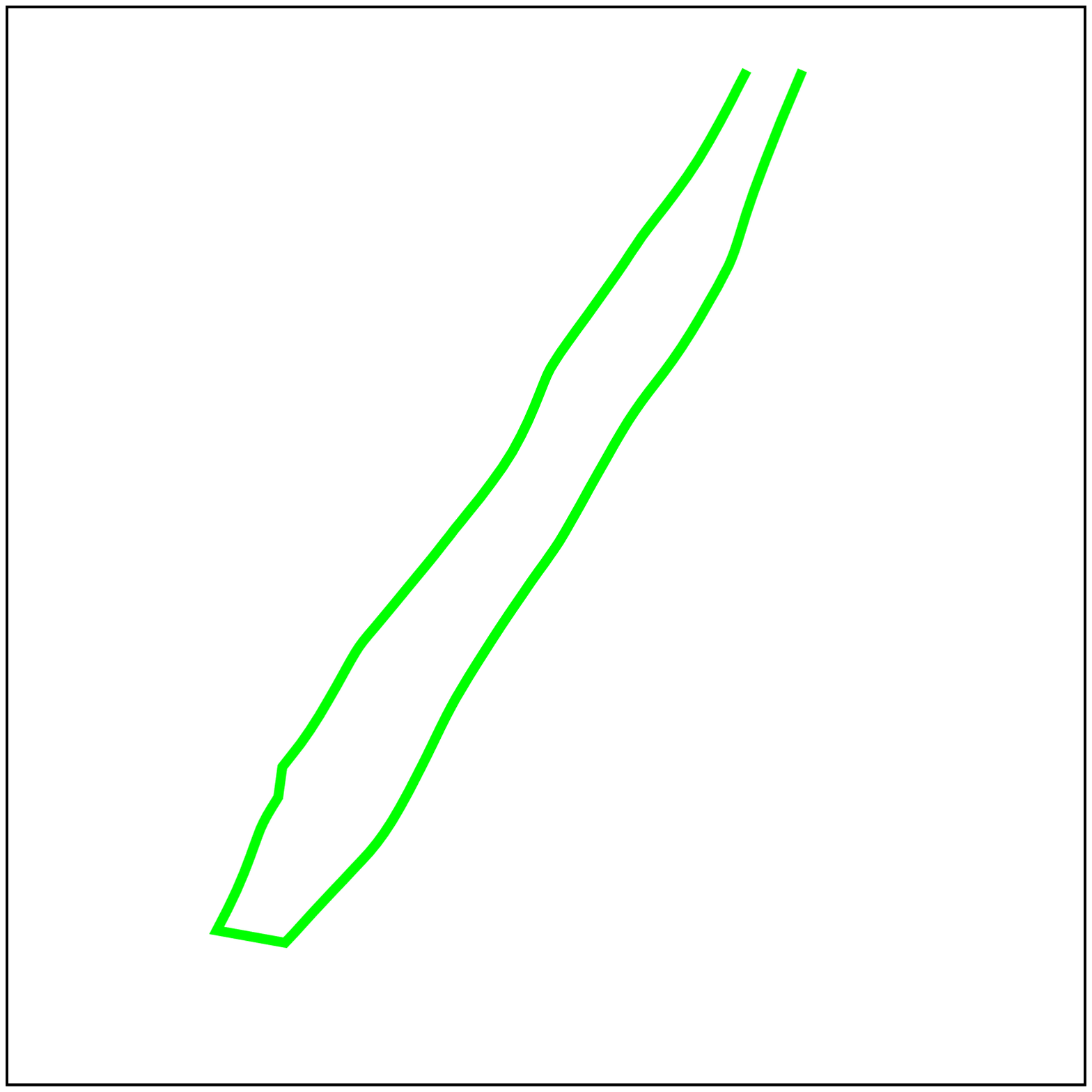}}
\showPur{\put(0,170){\includegraphics[angle=270,width=8cm]{valAfblo.eps}}}
\end{picture}}
\put(270,200){\begin{picture}(100,202)(55,0)
\put(-55,202){\includegraphics[angle=270,width=11.8cm]{mlSigLHC8.eps}}
\put(0,170){\includegraphics[angle=270,width=8cm]{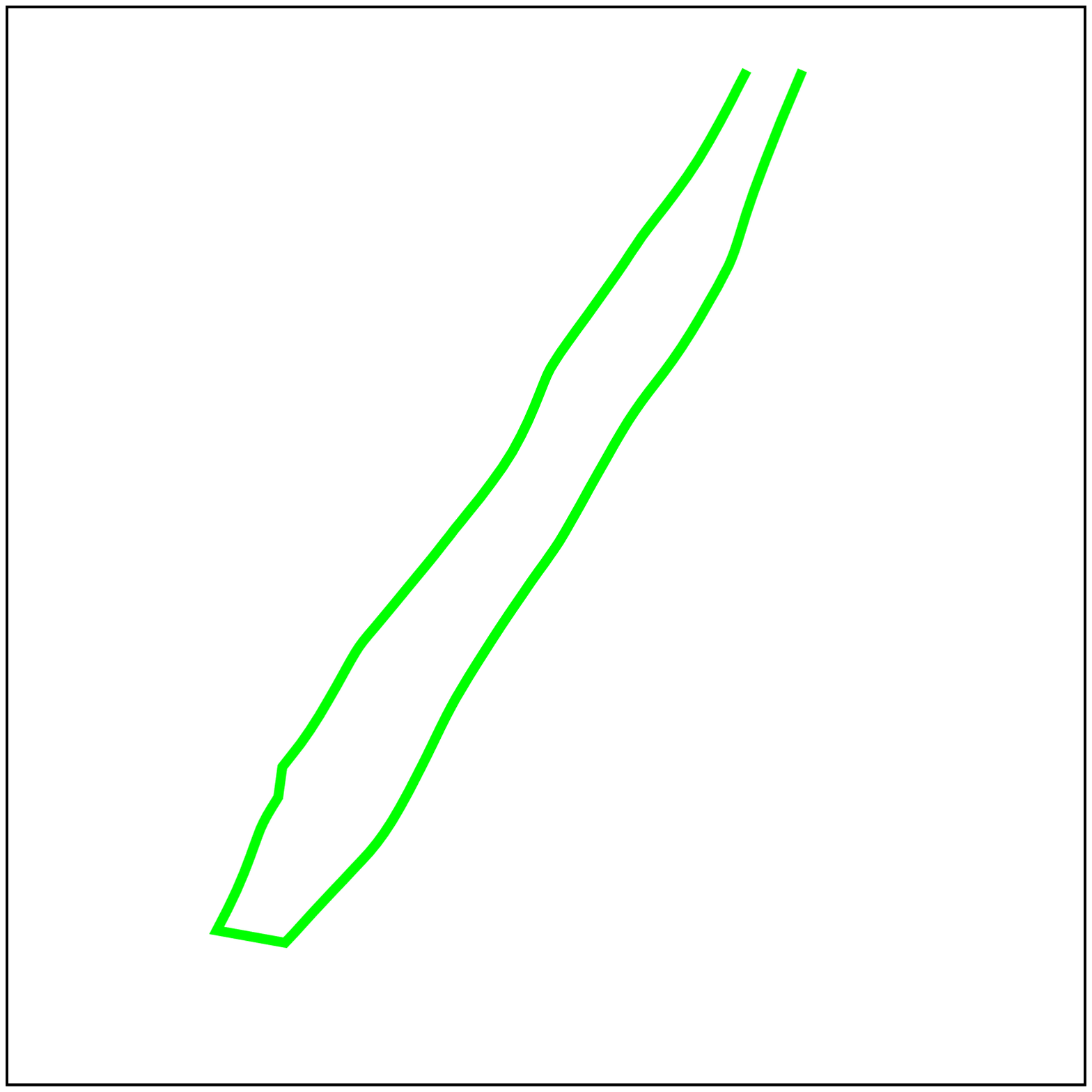}}
\showPur{\put(0,170){\includegraphics[angle=270,width=8cm]{valSigLhc8.eps}}}
\end{picture}
}
\put(30,0){\begin{picture}(100,202)(55,0)
\put(-55,202){\includegraphics[angle=270,width=11.8cm]{mlAfbhigh.eps}}
\put(0,170){\includegraphics[angle=270,width=8cm]{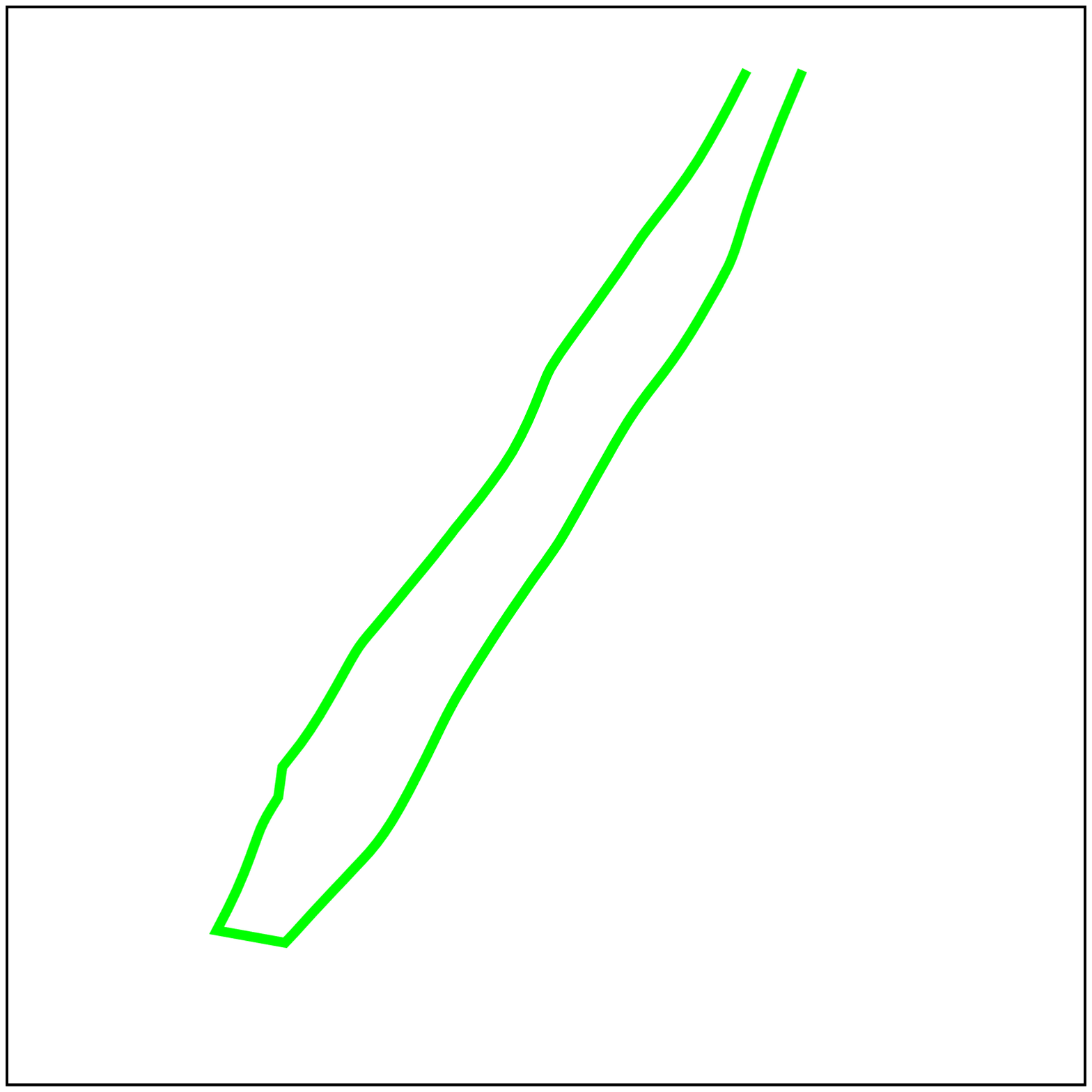}}
\showPur{\put(0,170){\includegraphics[angle=270,width=8cm]{valAfbhi.eps}}}
\end{picture}}
\put(270,0){\begin{picture}(100,202)(55,0)
\put(-55,202){\includegraphics[angle=270,width=11.8cm]{mlSigLHC.eps}}
\put(0,170){\includegraphics[angle=270,width=8cm]{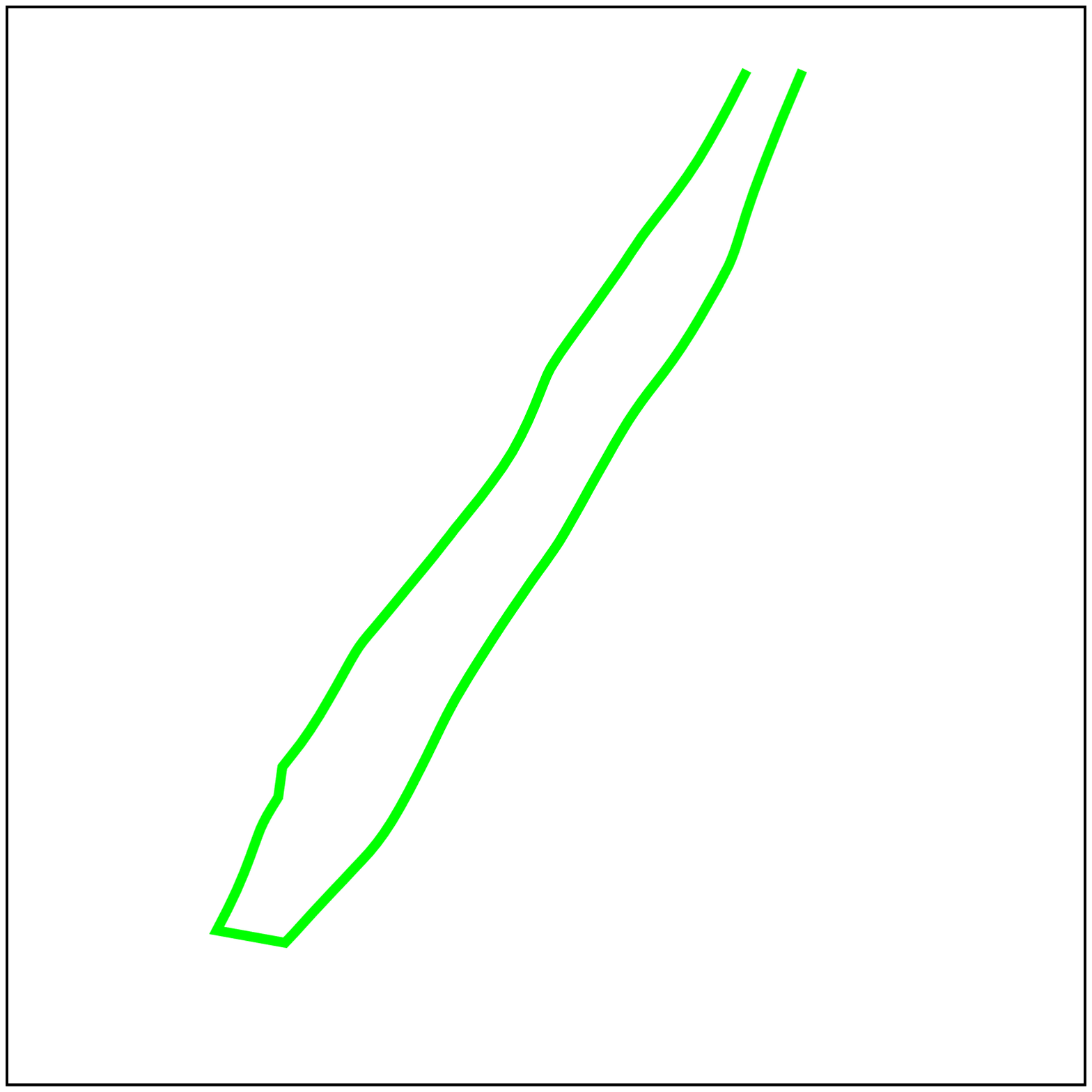}}
\showPur{\put(0,170){\includegraphics[angle=270,width=8cm]{valSigLhc.eps}}}
\end{picture}
}
    \put(0,370){(a)}
    \put(230,370){(b)}
    \put(0,175){(c)}
    \put(230,175){(d)}
\end{picture}
\caption{New physics contribution  to (a) the Tevatron forward backward
  asymmetry with $m_{t \bar t}<450$ GeV, (b)
  8 TeV LHC $t \bar t$ 
  production cross-section, (c) the Tevatron forward backward
  asymmetry with $m_{t \bar t}>450$ GeV, (d) 7  TeV LHC $t \bar t$ 
  production cross-section. Inside the
  contours, the bounds in Table~\protect\ref{tab:consts} are respected.
\label{fig:afbsig}}
\end{center}
\end{figure*}
}

Fig.~\ref{fig:afb} shows the effect of the constraints upon the parameter
space of the model. Encouragingly, there is some parameter space which
simultaneously fits
all of the relevant current $t \bar t$ constraints, shown by the region
enclosed by the solid contour.
By drawing additional contours which we do not show here
for reasons of clarity, 
we find the values of the various observables which are 
within the good fit region. 
These predictions are displayed in Table~\ref{tab:preds} and should aid future
tests of the model. 
ATLAS obtained a statistical error on the charge asymmetry of 0.028, and a
systematic error of 0.024 in 1 fb${-1}$ of 7 TeV LHC
data~\cite{ATLAS:2012an}. We expect both errors to scale like
$1/\sqrt{\mathcal{L}}$, since the systematics are also measurement dominated.
Thus we expect, for 10$^{-1}$fb, that each error will be around 0.01, which
will certainly test most of the parameter space, where $\Delta A_C^y$ is
higher.  

On the day of completion of this article, Ref.~\cite{kk} appeared on the
electronic archive. It considers the effect of the exchange of a charge $-1/3$
scalar
triplet diquark, and the bounds on parameter space coming from $A_{FB}$,
$A_C$, $\sigma^{TEV}_{t \bar   t}$ and
$\sigma^{LHC7}_{t \bar t}$ measurements agree with those shown here and aside
from the fact that the paper also uses {\tt MadGraph} to do the simulations,
it provides an independent confirmation of some of our results. 

All of the   predictions above apply to a generic model containing a colour
triplet scalar of charge $-1/3$ coupling to $d_R t_R$. 
Such models are extremely strongly constrained by atomic parity
violation (APV) constraints~\cite{Gresham:2012wc}. The RPV
MSSM, 
however, has many additional particles and interactions and the possibility
for cancellations of different contributions to the APV is open. 
Indeed,
a very recent paper~\cite{Dupuis:2012is}
has shown that stop mixing contributions to APV can completely cancel the
contribution from $\lambda''_{313} \neq 0$. 

It was recently
argued that baryon number 
violating couplings such as $\lambda''_{313}$ and a light stop 
could simultaneously explain the naturalness of SUSY and its evasion of 7 TeV LHC
searches based on large missing transverse momentum~\cite{Allanach:2012vj}.
SUSY
implies additional interactions with identical coupling strengths: in
particular,   ${\mathcal L}=(\lambda''_{313})^* {\tilde t}_R d_R^T C^\dag
b_R+H.c.+\ldots$, 
where $C$ is
the charge conjugation matrix and $^T$ denotes transpose. 
If ${\tilde t}_R$ is not too heavy, $t-$channel stop exchange should 
induce a $b \bar b$ forward backward asymmetry~\cite{Bai:2011ed}.
One may also expect resonant stop production to
produce a bump in the $m_{jj}$ distribution in two jet events where one of the
jets is a $b$. However, such a signal will be suppressed by small bottom
PDFs and further study is required to establish
the viability of detection.
\begin{table}
\begin{tabular}{c|c} \hline
$0.037<\Delta A_{FB}<0.11$ & $0.017<\Delta A_C^y<0.045$ \\
$-0.1<\Delta \sigma^{TEV}_{t \bar t}/\mbox{pb} <1.8 $ &
$25<\Delta \sigma^{TEV}_{t \bar t}(\mbox{bin})$/fb$<76$ \\
$0.006 < \Delta A_{FB}^l < 0.07$ & $0.062 < \Delta A_{FB}^h < 0.16$ \\
$8<\Delta \sigma_{t \bar t}^{LHC7}/\mbox{pb} < 25$ & 
$13<\Delta \sigma_{t \bar t}^{LHC8}/\mbox{pb} < 33$ \\
\hline
\end{tabular}
\caption{Predicted values of new physics contribution to various observables
  for parameter space that passes all current experimental constraints. 
\label{tab:preds}}
\end{table}

In summary, we have shown that the RPV MSSM can explain
the anomalously large $A_{FB}$
measured at the Tevatron experiments, provided that couplings are chosen in a
region that may violate perturbativity below the GUT scale. 
The $t-$channel exchange of a right-handed sbottom of
mass 300-1200 GeV which couples to $d_R t_R$ with an interaction strength
bigger than $2.2$ is required. 
There is parameter space which passes all
relevant constraints from the Tevatron and the LHC\@. 
Future LHC signals are predicted to be: a small positive asymmetry parameter
$0.017<\Delta A_C^y<0.045$ and a $t \bar t$ production cross-section that
should be larger than the SM prediction by at least 8 (13) pb at 7 (8)
TeV. Assuming that the uncertainties $\propto 1/\sqrt{{\mathcal L}}$ (where
${\mathcal L}$ is the integrated luminosity), there
should be 
enough information in 10 fb$^{-1}$ of 7 TeV LHC data to exclude such an
enhancement of the production cross-section to the 95$\%$ confidence level.  
Further confirmation through the top charge asymmetry may take longer, but is
within reach of 20 fb$^{-1}$.
These predictions, along with the other predictions in
Table~\ref{tab:preds}, provide targets for ongoing LHC tests of the model. 
Top spin observables also provide an interesting way to discriminate models of
new physics that explain the Tevatron measurements
$A_{FB}$~\cite{Berger:2012nw,Fajfer:2012si}, and it 
would be an interesting future project to investigate the predictions for them
coming from $\lambda''_{313}$.
If the sbottoms are not too heavy, then there is the possibility of
confirming the mechanism through ${\tilde
  b}_R {\tilde b}_R^*$ production at the LHC, which would then
decay to $t \bar t$ plus 2 jets, with an invariant mass bump in the combined 
$t$ and jet mass. A discovery such as this would provide a definitive test of
the model. We leave its investigation to future work. 

\section*{Acknowledgements}
This work has been partially supported by STFC\@. BCA thanks other
members of the Cambridge SUSY Working group for helpful suggestions
(particularly B Webber for help with numerical checks) and TIFR
for support under the adjunct faculty scheme and IPPP for support under the
associateship scheme.

\bibliographystyle{apsrev4-1}
\bibliography{a}

\begin{thebibliography}{37}%
\makeatletter
\providecommand \@ifxundefined [1]{%
 \@ifx{#1\undefined}
}%
\providecommand \@ifnum [1]{%
 \ifnum #1\expandafter \@firstoftwo
 \else \expandafter \@secondoftwo
 \fi
}%
\providecommand \@ifx [1]{%
 \ifx #1\expandafter \@firstoftwo
 \else \expandafter \@secondoftwo
 \fi
}%
\providecommand \natexlab [1]{#1}%
\providecommand \enquote  [1]{``#1''}%
\providecommand \bibnamefont  [1]{#1}%
\providecommand \bibfnamefont [1]{#1}%
\providecommand \citenamefont [1]{#1}%
\providecommand \href@noop [0]{\@secondoftwo}%
\providecommand \href [0]{\begingroup \@sanitize@url \@href}%
\providecommand \@href[1]{\@@startlink{#1}\@@href}%
\providecommand \@@href[1]{\endgroup#1\@@endlink}%
\providecommand \@sanitize@url [0]{\catcode `\\12\catcode `\$12\catcode
  `\&12\catcode `\#12\catcode `\^12\catcode `\_12\catcode `\%12\relax}%
\providecommand \@@startlink[1]{}%
\providecommand \@@endlink[0]{}%
\providecommand \url  [0]{\begingroup\@sanitize@url \@url }%
\providecommand \@url [1]{\endgroup\@href {#1}{\urlprefix }}%
\providecommand \urlprefix  [0]{URL }%
\providecommand \Eprint [0]{\href }%
\providecommand \doibase [0]{http://dx.doi.org/}%
\providecommand \selectlanguage [0]{\@gobble}%
\providecommand \bibinfo  [0]{\@secondoftwo}%
\providecommand \bibfield  [0]{\@secondoftwo}%
\providecommand \translation [1]{[#1]}%
\providecommand \BibitemOpen [0]{}%
\providecommand \bibitemStop [0]{}%
\providecommand \bibitemNoStop [0]{.\EOS\space}%
\providecommand \EOS [0]{\spacefactor3000\relax}%
\providecommand \BibitemShut  [1]{\csname bibitem#1\endcsname}%
\let\auto@bib@innerbib\@empty
\bibitem [{\citenamefont {Ahrens}\ \emph {et~al.}(2011)\citenamefont {Ahrens},
  \citenamefont {Ferroglia}, \citenamefont {Neubert}, \citenamefont {Pecjak},\
  and\ \citenamefont {Yang}}]{Ahrens:2011uf}%
  \BibitemOpen
  \bibfield  {author} {\bibinfo {author} {\bibfnamefont {V.}~\bibnamefont
  {Ahrens}}, \bibinfo {author} {\bibfnamefont {A.}~\bibnamefont {Ferroglia}},
  \bibinfo {author} {\bibfnamefont {M.}~\bibnamefont {Neubert}}, \bibinfo
  {author} {\bibfnamefont {B.~D.}\ \bibnamefont {Pecjak}}, \ and\ \bibinfo
  {author} {\bibfnamefont {L.~L.}\ \bibnamefont {Yang}},\ }\href@noop {}
  {\bibfield  {journal} {\bibinfo  {journal} {Phys.Rev.}\ }\textbf {\bibinfo
  {volume} {D84}},\ \bibinfo {pages} {074004} (\bibinfo {year} {2011})},\
  \Eprint {http://arxiv.org/abs/1106.6051} {arXiv:1106.6051 [hep-ph]}
  \BibitemShut {NoStop}%
\bibitem [{\citenamefont {Skands}\ \emph {et~al.}(2012)\citenamefont {Skands},
  \citenamefont {Webber},\ and\ \citenamefont {Winter}}]{Skands:2012mm}%
  \BibitemOpen
  \bibfield  {author} {\bibinfo {author} {\bibfnamefont {P.~Z.}\ \bibnamefont
  {Skands}}, \bibinfo {author} {\bibfnamefont {B.~R.}\ \bibnamefont {Webber}},
  \ and\ \bibinfo {author} {\bibfnamefont {J.}~\bibnamefont {Winter}},\
  }\href@noop {} {\  (\bibinfo {year} {2012})},\ \Eprint
  {http://arxiv.org/abs/1205.1466} {arXiv:1205.1466 [hep-ph]} \BibitemShut
  {NoStop}%
\bibitem [{\citenamefont {Aaltonen}\ \emph {et~al.}(2011)\citenamefont
  {Aaltonen} \emph {et~al.}}]{PhysRevD.83.112003}%
  \BibitemOpen
  \bibfield  {author} {\bibinfo {author} {\bibfnamefont {T.}~\bibnamefont
  {Aaltonen}} \emph {et~al.} (\bibinfo {collaboration} {CDF Collaboration}),\
  }\href {\doibase 10.1103/PhysRevD.83.112003} {\bibfield  {journal} {\bibinfo
  {journal} {Phys. Rev. D}\ }\textbf {\bibinfo {volume} {83}},\ \bibinfo
  {pages} {112003} (\bibinfo {year} {2011})}\BibitemShut {NoStop}%
\bibitem [{\citenamefont {Aaltonen}\ \emph {et~al.}(2012)\citenamefont
  {Aaltonen} \emph {et~al.}}]{10584}%
  \BibitemOpen
  \bibfield  {author} {\bibinfo {author} {\bibfnamefont {T.}~\bibnamefont
  {Aaltonen}} \emph {et~al.} (\bibinfo {collaboration} {CDF Collaboration}),\
  }\href@noop {} {} (\bibinfo {year} {2012}),\ \bibinfo {note} {cDF note
  10584}\BibitemShut {NoStop}%
\bibitem [{\citenamefont {Abazov}\ \emph {et~al.}(2011)\citenamefont {Abazov}
  \emph {et~al.}}]{PhysRevD.84.112005}%
  \BibitemOpen
  \bibfield  {author} {\bibinfo {author} {\bibfnamefont {V.~M.}\ \bibnamefont
  {Abazov}} \emph {et~al.} (\bibinfo {collaboration} {The D0 Collaboration}),\
  }\href {\doibase 10.1103/PhysRevD.84.112005} {\bibfield  {journal} {\bibinfo
  {journal} {Phys. Rev. D}\ }\textbf {\bibinfo {volume} {84}},\ \bibinfo
  {pages} {112005} (\bibinfo {year} {2011})}\BibitemShut {NoStop}%
\bibitem [{\citenamefont {Aad}\ \emph {et~al.}(2012)\citenamefont {Aad} \emph
  {et~al.}}]{ATLAS:2012an}%
  \BibitemOpen
  \bibfield  {author} {\bibinfo {author} {\bibfnamefont {G.}~\bibnamefont
  {Aad}} \emph {et~al.} (\bibinfo {collaboration} {ATLAS Collaboration}),\
  }\href@noop {} {\  (\bibinfo {year} {2012})},\ \Eprint
  {http://arxiv.org/abs/1203.4211} {arXiv:1203.4211 [hep-ex]} \BibitemShut
  {NoStop}%
\bibitem [{\citenamefont {Chatrchyan}\ \emph {et~al.}(2012)\citenamefont
  {Chatrchyan} \emph {et~al.}}]{Chatrchyan:2011hk}%
  \BibitemOpen
  \bibfield  {author} {\bibinfo {author} {\bibfnamefont {S.}~\bibnamefont
  {Chatrchyan}} \emph {et~al.} (\bibinfo {collaboration} {CMS Collaboration}),\
  }\href@noop {} {\bibfield  {journal} {\bibinfo  {journal} {Phys.Lett.}\
  }\textbf {\bibinfo {volume} {B709}},\ \bibinfo {pages} {28} (\bibinfo {year}
  {2012})},\ \Eprint {http://arxiv.org/abs/1112.5100} {arXiv:1112.5100
  [hep-ex]} \BibitemShut {NoStop}%
\bibitem [{\citenamefont {Shu}\ \emph {et~al.}(2010)\citenamefont {Shu},
  \citenamefont {Tait},\ and\ \citenamefont {Wang}}]{Shu:2009xf}%
  \BibitemOpen
  \bibfield  {author} {\bibinfo {author} {\bibfnamefont {J.}~\bibnamefont
  {Shu}}, \bibinfo {author} {\bibfnamefont {T.~M.}\ \bibnamefont {Tait}}, \
  and\ \bibinfo {author} {\bibfnamefont {K.}~\bibnamefont {Wang}},\ }\href
  {\doibase 10.1103/PhysRevD.81.034012} {\bibfield  {journal} {\bibinfo
  {journal} {Phys.Rev.}\ }\textbf {\bibinfo {volume} {D81}},\ \bibinfo {pages}
  {034012} (\bibinfo {year} {2010})},\ \Eprint {http://arxiv.org/abs/0911.3237}
  {arXiv:0911.3237 [hep-ph]} \BibitemShut {NoStop}%
\bibitem [{\citenamefont {Aguilar-Saavedra}(2012)}]{AguilarSaavedra:2012ma}%
  \BibitemOpen
  \bibfield  {author} {\bibinfo {author} {\bibfnamefont {J.}~\bibnamefont
  {Aguilar-Saavedra}},\ }\href@noop {} {\  (\bibinfo {year} {2012})},\ \Eprint
  {http://arxiv.org/abs/1202.2382} {arXiv:1202.2382 [hep-ph]} \BibitemShut
  {NoStop}%
\bibitem [{\citenamefont {Blum}\ \emph {et~al.}(2011)\citenamefont {Blum},
  \citenamefont {Hochberg},\ and\ \citenamefont {Nir}}]{Blum:2011fa}%
  \BibitemOpen
  \bibfield  {author} {\bibinfo {author} {\bibfnamefont {K.}~\bibnamefont
  {Blum}}, \bibinfo {author} {\bibfnamefont {Y.}~\bibnamefont {Hochberg}}, \
  and\ \bibinfo {author} {\bibfnamefont {Y.}~\bibnamefont {Nir}},\ }\href@noop
  {} {\bibfield  {journal} {\bibinfo  {journal} {JHEP}\ }\textbf {\bibinfo
  {volume} {1110}},\ \bibinfo {pages} {124} (\bibinfo {year} {2011})},\ \Eprint
  {http://arxiv.org/abs/1107.4350} {arXiv:1107.4350 [hep-ph]} \BibitemShut
  {NoStop}%
\bibitem [{\citenamefont {Cao}\ \emph {et~al.}(2010)\citenamefont {Cao},
  \citenamefont {Heng}, \citenamefont {Wu},\ and\ \citenamefont
  {Yang}}]{Cao:2009uz}%
  \BibitemOpen
  \bibfield  {author} {\bibinfo {author} {\bibfnamefont {J.}~\bibnamefont
  {Cao}}, \bibinfo {author} {\bibfnamefont {Z.}~\bibnamefont {Heng}}, \bibinfo
  {author} {\bibfnamefont {L.}~\bibnamefont {Wu}}, \ and\ \bibinfo {author}
  {\bibfnamefont {J.~M.}\ \bibnamefont {Yang}},\ }\href {\doibase
  10.1103/PhysRevD.81.014016} {\bibfield  {journal} {\bibinfo  {journal}
  {Phys.Rev.}\ }\textbf {\bibinfo {volume} {D81}},\ \bibinfo {pages} {014016}
  (\bibinfo {year} {2010})},\ \Eprint {http://arxiv.org/abs/0912.1447}
  {arXiv:0912.1447 [hep-ph]} \BibitemShut {NoStop}%
\bibitem [{\citenamefont {Ghosh}\ \emph {et~al.}(1997)\citenamefont {Ghosh},
  \citenamefont {Raychaudhuri},\ and\ \citenamefont {Sridhar}}]{Ghosh:1996bm}%
  \BibitemOpen
  \bibfield  {author} {\bibinfo {author} {\bibfnamefont {D.~K.}\ \bibnamefont
  {Ghosh}}, \bibinfo {author} {\bibfnamefont {S.}~\bibnamefont {Raychaudhuri}},
  \ and\ \bibinfo {author} {\bibfnamefont {K.}~\bibnamefont {Sridhar}},\ }\href
  {\doibase 10.1016/S0370-2693(97)00117-2} {\bibfield  {journal} {\bibinfo
  {journal} {Phys.Lett.}\ }\textbf {\bibinfo {volume} {B396}},\ \bibinfo
  {pages} {177} (\bibinfo {year} {1997})},\ \Eprint
  {http://arxiv.org/abs/hep-ph/9608352} {arXiv:hep-ph/9608352 [hep-ph]}
  \BibitemShut {NoStop}%
\bibitem [{\citenamefont {Dupuis}\ and\ \citenamefont
  {Cline}(2012)}]{Dupuis:2012is}%
  \BibitemOpen
  \bibfield  {author} {\bibinfo {author} {\bibfnamefont {G.}~\bibnamefont
  {Dupuis}}\ and\ \bibinfo {author} {\bibfnamefont {J.~M.}\ \bibnamefont
  {Cline}},\ }\href@noop {} {\  (\bibinfo {year} {2012})},\ \Eprint
  {http://arxiv.org/abs/1206.1845} {arXiv:1206.1845 [hep-ph]} \BibitemShut
  {NoStop}%
\bibitem [{\citenamefont {Alwall}\ \emph {et~al.}(2011)\citenamefont {Alwall},
  \citenamefont {Herquet}, \citenamefont {Maltoni}, \citenamefont {Mattelaer},\
  and\ \citenamefont {Stelzer}}]{Alwall:2011uj}%
  \BibitemOpen
  \bibfield  {author} {\bibinfo {author} {\bibfnamefont {J.}~\bibnamefont
  {Alwall}}, \bibinfo {author} {\bibfnamefont {M.}~\bibnamefont {Herquet}},
  \bibinfo {author} {\bibfnamefont {F.}~\bibnamefont {Maltoni}}, \bibinfo
  {author} {\bibfnamefont {O.}~\bibnamefont {Mattelaer}}, \ and\ \bibinfo
  {author} {\bibfnamefont {T.}~\bibnamefont {Stelzer}},\ }\href {\doibase
  10.1007/JHEP06(2011)128} {\bibfield  {journal} {\bibinfo  {journal} {JHEP}\
  }\textbf {\bibinfo {volume} {1106}},\ \bibinfo {pages} {128} (\bibinfo {year}
  {2011})},\ \Eprint {http://arxiv.org/abs/1106.0522} {arXiv:1106.0522
  [hep-ph]} \BibitemShut {NoStop}%
\bibitem [{\citenamefont {Leone}()}]{moriond}%
  \BibitemOpen
  \bibfield  {author} {\bibinfo {author} {\bibfnamefont {S.}~\bibnamefont
  {Leone}} (\bibinfo {collaboration} {CDF}),\ }\href
  {http://indico.in2p3.fr/getFile.py/access?contribId=22&sessionId=9&resId=0&m%
aterialId=slides&confId=6001} {\enquote {\bibinfo {title} {Top quark production
  at the tevatron},}\ }\bibinfo {note} {Presented at Electroweak Session of
  Rencontres de Moriond 2012}\BibitemShut {NoStop}%
\bibitem [{\citenamefont {Allanach}\ \emph {et~al.}(1999)\citenamefont
  {Allanach}, \citenamefont {Dedes},\ and\ \citenamefont
  {Dreiner}}]{Allanach:1999ic}%
  \BibitemOpen
  \bibfield  {author} {\bibinfo {author} {\bibfnamefont {B.}~\bibnamefont
  {Allanach}}, \bibinfo {author} {\bibfnamefont {A.}~\bibnamefont {Dedes}}, \
  and\ \bibinfo {author} {\bibfnamefont {H.~K.}\ \bibnamefont {Dreiner}},\
  }\href {\doibase 10.1103/PhysRevD.60.075014} {\bibfield  {journal} {\bibinfo
  {journal} {Phys.Rev.}\ }\textbf {\bibinfo {volume} {D60}},\ \bibinfo {pages}
  {075014} (\bibinfo {year} {1999})},\ \Eprint
  {http://arxiv.org/abs/hep-ph/9906209} {arXiv:hep-ph/9906209 [hep-ph]}
  \BibitemShut {NoStop}%
\bibitem [{\citenamefont {Barbier}\ \emph {et~al.}(2005)\citenamefont
  {Barbier}, \citenamefont {Berat}, \citenamefont {Besancon}, \citenamefont
  {Chemtob}, \citenamefont {Deandrea} \emph {et~al.}}]{Barbier:2004ez}%
  \BibitemOpen
  \bibfield  {author} {\bibinfo {author} {\bibfnamefont {R.}~\bibnamefont
  {Barbier}}, \bibinfo {author} {\bibfnamefont {C.}~\bibnamefont {Berat}},
  \bibinfo {author} {\bibfnamefont {M.}~\bibnamefont {Besancon}}, \bibinfo
  {author} {\bibfnamefont {M.}~\bibnamefont {Chemtob}}, \bibinfo {author}
  {\bibfnamefont {A.}~\bibnamefont {Deandrea}},  \emph {et~al.},\ }\href
  {\doibase 10.1016/j.physrep.2005.08.006} {\bibfield  {journal} {\bibinfo
  {journal} {Phys.Rept.}\ }\textbf {\bibinfo {volume} {420}},\ \bibinfo {pages}
  {1} (\bibinfo {year} {2005})},\ \Eprint {http://arxiv.org/abs/hep-ph/0406039}
  {arXiv:hep-ph/0406039 [hep-ph]} \BibitemShut {NoStop}%
\bibitem [{\citenamefont {Chang}\ and\ \citenamefont
  {Keung}(1996)}]{Chang:1996sw}%
  \BibitemOpen
  \bibfield  {author} {\bibinfo {author} {\bibfnamefont {D.}~\bibnamefont
  {Chang}}\ and\ \bibinfo {author} {\bibfnamefont {W.-Y.}\ \bibnamefont
  {Keung}},\ }\href {\doibase 10.1016/S0370-2693(96)01271-3} {\bibfield
  {journal} {\bibinfo  {journal} {Phys.Lett.}\ }\textbf {\bibinfo {volume}
  {B389}},\ \bibinfo {pages} {294} (\bibinfo {year} {1996})},\ \Eprint
  {http://arxiv.org/abs/hep-ph/9608313} {arXiv:hep-ph/9608313 [hep-ph]}
  \BibitemShut {NoStop}%
\bibitem [{\citenamefont {Bhattacharyya}\ \emph {et~al.}(1995)\citenamefont
  {Bhattacharyya}, \citenamefont {Choudhury},\ and\ \citenamefont
  {Sridhar}}]{Bhattacharyya:1995bw}%
  \BibitemOpen
  \bibfield  {author} {\bibinfo {author} {\bibfnamefont {G.}~\bibnamefont
  {Bhattacharyya}}, \bibinfo {author} {\bibfnamefont {D.}~\bibnamefont
  {Choudhury}}, \ and\ \bibinfo {author} {\bibfnamefont {K.}~\bibnamefont
  {Sridhar}},\ }\href {\doibase 10.1016/0370-2693(95)00701-L} {\bibfield
  {journal} {\bibinfo  {journal} {Phys.Lett.}\ }\textbf {\bibinfo {volume}
  {B355}},\ \bibinfo {pages} {193} (\bibinfo {year} {1995})},\ \Eprint
  {http://arxiv.org/abs/hep-ph/9504314} {arXiv:hep-ph/9504314 [hep-ph]}
  \BibitemShut {NoStop}%
\bibitem [{\citenamefont {Giudice}\ \emph {et~al.}(2011)\citenamefont
  {Giudice}, \citenamefont {Gripaios},\ and\ \citenamefont
  {Sundrum}}]{Giudice:2011ak}%
  \BibitemOpen
  \bibfield  {author} {\bibinfo {author} {\bibfnamefont {G.~F.}\ \bibnamefont
  {Giudice}}, \bibinfo {author} {\bibfnamefont {B.}~\bibnamefont {Gripaios}}, \
  and\ \bibinfo {author} {\bibfnamefont {R.}~\bibnamefont {Sundrum}},\ }\href
  {\doibase 10.1007/JHEP08(2011)055} {\bibfield  {journal} {\bibinfo  {journal}
  {JHEP}\ }\textbf {\bibinfo {volume} {1108}},\ \bibinfo {pages} {055}
  (\bibinfo {year} {2011})},\ \Eprint {http://arxiv.org/abs/1105.3161}
  {arXiv:1105.3161 [hep-ph]} \BibitemShut {NoStop}%
\bibitem [{\citenamefont {Slavich}(2001)}]{Slavich:2000xm}%
  \BibitemOpen
  \bibfield  {author} {\bibinfo {author} {\bibfnamefont {P.}~\bibnamefont
  {Slavich}},\ }\href {\doibase 10.1016/S0550-3213(00)00700-8} {\bibfield
  {journal} {\bibinfo  {journal} {Nucl.Phys.}\ }\textbf {\bibinfo {volume}
  {B595}},\ \bibinfo {pages} {33} (\bibinfo {year} {2001})},\ \Eprint
  {http://arxiv.org/abs/hep-ph/0008270} {arXiv:hep-ph/0008270 [hep-ph]}
  \BibitemShut {NoStop}%
\bibitem [{\citenamefont {Aaltonen}\ \emph
  {et~al.}(2009{\natexlab{a}})\citenamefont {Aaltonen} \emph {et~al.}}]{9913}%
  \BibitemOpen
  \bibfield  {author} {\bibinfo {author} {\bibfnamefont {T.}~\bibnamefont
  {Aaltonen}} \emph {et~al.} (\bibinfo {collaboration} {CDF Collaboration}),\
  }\href@noop {} {} (\bibinfo {year} {2009}{\natexlab{a}}),\ \bibinfo {note}
  {{CDF} note 9913}\BibitemShut {NoStop}%
\bibitem [{\citenamefont {Aaltonen}\ \emph
  {et~al.}(2009{\natexlab{b}})\citenamefont {Aaltonen} \emph
  {et~al.}}]{Aaltonen:2009iz}%
  \BibitemOpen
  \bibfield  {author} {\bibinfo {author} {\bibfnamefont {T.}~\bibnamefont
  {Aaltonen}} \emph {et~al.} (\bibinfo {collaboration} {CDF Collaboration}),\
  }\href {\doibase 10.1103/PhysRevLett.102.222003} {\bibfield  {journal}
  {\bibinfo  {journal} {Phys.Rev.Lett.}\ }\textbf {\bibinfo {volume} {102}},\
  \bibinfo {pages} {222003} (\bibinfo {year} {2009}{\natexlab{b}})},\ \Eprint
  {http://arxiv.org/abs/0903.2850} {arXiv:0903.2850 [hep-ex]} \BibitemShut
  {NoStop}%
\bibitem [{\citenamefont {Almeida}\ \emph {et~al.}(2008)\citenamefont
  {Almeida}, \citenamefont {Sterman},\ and\ \citenamefont
  {Vogelsang}}]{Almeida:2008ug}%
  \BibitemOpen
  \bibfield  {author} {\bibinfo {author} {\bibfnamefont {L.~G.}\ \bibnamefont
  {Almeida}}, \bibinfo {author} {\bibfnamefont {G.~F.}\ \bibnamefont
  {Sterman}}, \ and\ \bibinfo {author} {\bibfnamefont {W.}~\bibnamefont
  {Vogelsang}},\ }\href {\doibase 10.1103/PhysRevD.78.014008} {\bibfield
  {journal} {\bibinfo  {journal} {Phys.Rev.}\ }\textbf {\bibinfo {volume}
  {D78}},\ \bibinfo {pages} {014008} (\bibinfo {year} {2008})},\ \Eprint
  {http://arxiv.org/abs/0805.1885} {arXiv:0805.1885 [hep-ph]} \BibitemShut
  {NoStop}%
\bibitem [{\citenamefont {Ahrens}\ \emph {et~al.}(2010)\citenamefont {Ahrens},
  \citenamefont {Ferroglia}, \citenamefont {Neubert}, \citenamefont {Pecjak},\
  and\ \citenamefont {Yang}}]{Ahrens:2010zv}%
  \BibitemOpen
  \bibfield  {author} {\bibinfo {author} {\bibfnamefont {V.}~\bibnamefont
  {Ahrens}}, \bibinfo {author} {\bibfnamefont {A.}~\bibnamefont {Ferroglia}},
  \bibinfo {author} {\bibfnamefont {M.}~\bibnamefont {Neubert}}, \bibinfo
  {author} {\bibfnamefont {B.~D.}\ \bibnamefont {Pecjak}}, \ and\ \bibinfo
  {author} {\bibfnamefont {L.~L.}\ \bibnamefont {Yang}},\ }\href {\doibase
  10.1007/JHEP09(2010)097} {\bibfield  {journal} {\bibinfo  {journal} {JHEP}\
  }\textbf {\bibinfo {volume} {1009}},\ \bibinfo {pages} {097} (\bibinfo {year}
  {2010})},\ \Eprint {http://arxiv.org/abs/1003.5827} {arXiv:1003.5827
  [hep-ph]} \BibitemShut {NoStop}%
\bibitem [{\citenamefont {Aad}\ \emph {et~al.}()\citenamefont {Aad} \emph
  {et~al.}}]{atlas1}%
  \BibitemOpen
  \bibfield  {author} {\bibinfo {author} {\bibfnamefont {G.}~\bibnamefont
  {Aad}} \emph {et~al.} (\bibinfo {collaboration} {ATLAS Collaboration}),\
  }\href@noop {} {\ }\bibinfo {note} {ATLAS-CONF-2011-121}\BibitemShut
  {NoStop}%
\bibitem [{\citenamefont {Chatrchyan}\ \emph {et~al.}()\citenamefont
  {Chatrchyan} \emph {et~al.}}]{cms1}%
  \BibitemOpen
  \bibfield  {author} {\bibinfo {author} {\bibfnamefont {S.}~\bibnamefont
  {Chatrchyan}} \emph {et~al.} (\bibinfo {collaboration} {CMS Collaboration}),\
  }\href@noop {} {\ }\bibinfo {note} {CMS PAS TOP-11-003}\BibitemShut {NoStop}%
\bibitem [{\citenamefont {Kretzer}\ \emph {et~al.}(2004)\citenamefont
  {Kretzer}, \citenamefont {Lai}, \citenamefont {Olness},\ and\ \citenamefont
  {Tung}}]{Kretzer:2003it}%
  \BibitemOpen
  \bibfield  {author} {\bibinfo {author} {\bibfnamefont {S.}~\bibnamefont
  {Kretzer}}, \bibinfo {author} {\bibfnamefont {H.}~\bibnamefont {Lai}},
  \bibinfo {author} {\bibfnamefont {F.}~\bibnamefont {Olness}}, \ and\ \bibinfo
  {author} {\bibfnamefont {W.}~\bibnamefont {Tung}},\ }\href {\doibase
  10.1103/PhysRevD.69.114005} {\bibfield  {journal} {\bibinfo  {journal}
  {Phys.Rev.}\ }\textbf {\bibinfo {volume} {D69}},\ \bibinfo {pages} {114005}
  (\bibinfo {year} {2004})},\ \Eprint {http://arxiv.org/abs/hep-ph/0307022}
  {arXiv:hep-ph/0307022 [hep-ph]} \BibitemShut {NoStop}%
\bibitem [{\citenamefont {Christensen}\ and\ \citenamefont
  {Duhr}(2009)}]{Christensen:2008py}%
  \BibitemOpen
  \bibfield  {author} {\bibinfo {author} {\bibfnamefont {N.~D.}\ \bibnamefont
  {Christensen}}\ and\ \bibinfo {author} {\bibfnamefont {C.}~\bibnamefont
  {Duhr}},\ }\href {\doibase 10.1016/j.cpc.2009.02.018} {\bibfield  {journal}
  {\bibinfo  {journal} {Comput.Phys.Commun.}\ }\textbf {\bibinfo {volume}
  {180}},\ \bibinfo {pages} {1614} (\bibinfo {year} {2009})},\ \Eprint
  {http://arxiv.org/abs/0806.4194} {arXiv:0806.4194 [hep-ph]} \BibitemShut
  {NoStop}%
\bibitem [{\citenamefont {Fuks}(2012)}]{Fuks:2012im}%
  \BibitemOpen
  \bibfield  {author} {\bibinfo {author} {\bibfnamefont {B.}~\bibnamefont
  {Fuks}},\ }\href {\doibase 10.1142/S0217751X12300074} {\bibfield  {journal}
  {\bibinfo  {journal} {Int.J.Mod.Phys.}\ }\textbf {\bibinfo {volume} {A27}},\
  \bibinfo {pages} {1230007} (\bibinfo {year} {2012})},\ \Eprint
  {http://arxiv.org/abs/1202.4769} {arXiv:1202.4769 [hep-ph]} \BibitemShut
  {NoStop}%
\bibitem [{\citenamefont {Duhr}\ and\ \citenamefont
  {Fuks}(2011)}]{Duhr:2011se}%
  \BibitemOpen
  \bibfield  {author} {\bibinfo {author} {\bibfnamefont {C.}~\bibnamefont
  {Duhr}}\ and\ \bibinfo {author} {\bibfnamefont {B.}~\bibnamefont {Fuks}},\
  }\href {\doibase 10.1016/j.cpc.2011.06.009} {\bibfield  {journal} {\bibinfo
  {journal} {Comput.Phys.Commun.}\ }\textbf {\bibinfo {volume} {182}},\
  \bibinfo {pages} {2404} (\bibinfo {year} {2011})},\ \Eprint
  {http://arxiv.org/abs/1102.4191} {arXiv:1102.4191 [hep-ph]} \BibitemShut
  {NoStop}%
\bibitem [{\citenamefont {Hagiwara}\ and\ \citenamefont {Nakamura}(2012)}]{kk}%
  \BibitemOpen
  \bibfield  {author} {\bibinfo {author} {\bibfnamefont {K.}~\bibnamefont
  {Hagiwara}}\ and\ \bibinfo {author} {\bibfnamefont {J.}~\bibnamefont
  {Nakamura}},\ }\href@noop {} {\  (\bibinfo {year} {2012})},\ \Eprint
  {http://arxiv.org/abs/1205.5005} {arXiv:1205.5005 [hep-ph]} \BibitemShut
  {NoStop}%
\bibitem [{\citenamefont {Gresham}\ \emph {et~al.}(2012)\citenamefont
  {Gresham}, \citenamefont {Kim}, \citenamefont {Tulin},\ and\ \citenamefont
  {Zurek}}]{Gresham:2012wc}%
  \BibitemOpen
  \bibfield  {author} {\bibinfo {author} {\bibfnamefont {M.~I.}\ \bibnamefont
  {Gresham}}, \bibinfo {author} {\bibfnamefont {I.-W.}\ \bibnamefont {Kim}},
  \bibinfo {author} {\bibfnamefont {S.}~\bibnamefont {Tulin}}, \ and\ \bibinfo
  {author} {\bibfnamefont {K.~M.}\ \bibnamefont {Zurek}},\ }\href@noop {} {\
  (\bibinfo {year} {2012})},\ \Eprint {http://arxiv.org/abs/1203.1320}
  {arXiv:1203.1320 [hep-ph]} \BibitemShut {NoStop}%
\bibitem [{\citenamefont {Allanach}\ and\ \citenamefont
  {Gripaios}(2012)}]{Allanach:2012vj}%
  \BibitemOpen
  \bibfield  {author} {\bibinfo {author} {\bibfnamefont {B.}~\bibnamefont
  {Allanach}}\ and\ \bibinfo {author} {\bibfnamefont {B.}~\bibnamefont
  {Gripaios}},\ }\href@noop {} {\  (\bibinfo {year} {2012})},\ \Eprint
  {http://arxiv.org/abs/1202.6616} {arXiv:1202.6616 [hep-ph]} \BibitemShut
  {NoStop}%
\bibitem [{\citenamefont {Bai}\ \emph {et~al.}(2011)\citenamefont {Bai},
  \citenamefont {Hewett}, \citenamefont {Kaplan},\ and\ \citenamefont
  {Rizzo}}]{Bai:2011ed}%
  \BibitemOpen
  \bibfield  {author} {\bibinfo {author} {\bibfnamefont {Y.}~\bibnamefont
  {Bai}}, \bibinfo {author} {\bibfnamefont {J.~L.}\ \bibnamefont {Hewett}},
  \bibinfo {author} {\bibfnamefont {J.}~\bibnamefont {Kaplan}}, \ and\ \bibinfo
  {author} {\bibfnamefont {T.~G.}\ \bibnamefont {Rizzo}},\ }\href {\doibase
  10.1007/JHEP03(2011)003} {\bibfield  {journal} {\bibinfo  {journal} {JHEP}\
  }\textbf {\bibinfo {volume} {1103}},\ \bibinfo {pages} {003} (\bibinfo {year}
  {2011})},\ \Eprint {http://arxiv.org/abs/1101.5203} {arXiv:1101.5203
  [hep-ph]} \BibitemShut {NoStop}%
\bibitem [{\citenamefont {Berger}\ \emph {et~al.}(2012)\citenamefont {Berger},
  \citenamefont {Cao}, \citenamefont {Chen}, \citenamefont {Yu},\ and\
  \citenamefont {Zhang}}]{Berger:2012nw}%
  \BibitemOpen
  \bibfield  {author} {\bibinfo {author} {\bibfnamefont {E.~L.}\ \bibnamefont
  {Berger}}, \bibinfo {author} {\bibfnamefont {Q.-H.}\ \bibnamefont {Cao}},
  \bibinfo {author} {\bibfnamefont {C.-R.}\ \bibnamefont {Chen}}, \bibinfo
  {author} {\bibfnamefont {J.-H.}\ \bibnamefont {Yu}}, \ and\ \bibinfo {author}
  {\bibfnamefont {H.}~\bibnamefont {Zhang}},\ }\href {\doibase
  10.1103/PhysRevLett.108.072002} {\bibfield  {journal} {\bibinfo  {journal}
  {Phys.Rev.Lett.}\ }\textbf {\bibinfo {volume} {108}},\ \bibinfo {pages}
  {072002} (\bibinfo {year} {2012})},\ \Eprint {http://arxiv.org/abs/1201.1790}
  {arXiv:1201.1790 [hep-ph]} \BibitemShut {NoStop}%
\bibitem [{\citenamefont {Fajfer}\ \emph {et~al.}(2012)\citenamefont {Fajfer},
  \citenamefont {Kamenik},\ and\ \citenamefont {Melic}}]{Fajfer:2012si}%
  \BibitemOpen
  \bibfield  {author} {\bibinfo {author} {\bibfnamefont {S.}~\bibnamefont
  {Fajfer}}, \bibinfo {author} {\bibfnamefont {J.~F.}\ \bibnamefont {Kamenik}},
  \ and\ \bibinfo {author} {\bibfnamefont {B.}~\bibnamefont {Melic}},\
  }\href@noop {} {\  (\bibinfo {year} {2012})},\ \Eprint
  {http://arxiv.org/abs/1205.0264} {arXiv:1205.0264 [hep-ph]} \BibitemShut
  {NoStop}%
\end{thebibliography}%

\end{document}